\PassOptionsToPackage{table,xcdraw}{xcolor}
\documentclass[journal]{IEEEtran}
\usepackage[table,xcdraw]{xcolor}
\usepackage{tikz}
\usepackage{tikzpagenodes}
\usepackage{tcolorbox}
\usepackage{fontawesome}
\tikzset{
  startstop/.style={rectangle, rounded corners, minimum width=3cm, minimum height=1cm, text centered, draw=black, fill=white!30},
  process/.style={rectangle, minimum width=3cm, minimum height=1cm, text centered, draw=black, fill=white!30},
  subprocess/.style={rectangle, minimum width=3cm, minimum height=1cm, text centered, draw=black, fill=white!30},
  arrow/.style={thick,->,>=stealth},
}
\def\BibTeX{{\rm B\kern-.05em{\sc i\kern-.025em b}\kern-.08em
    T\kern-.1667em\lower.7ex\hbox{E}\kern-.125emX}}

\usepackage{amssymb}
\usepackage[switch]{lineno}
\usepackage{hyperref}
\usepackage{color}
\usepackage{multirow}

\usepackage[normalem]{ulem}
\usepackage{xcolor}

\usepackage{booktabs} 
\usepackage{amssymb}
\usepackage{mathtools}
\usepackage{amsmath}
\usepackage{blkarray}
\usepackage{rotating} 
\usepackage[table,xcdraw]{xcolor}
\usepackage{makecell}
\usepackage{pifont}%
\usepackage{footmisc}
\usepackage{soul}
\usepackage{bbding}
\usepackage{enumerate}
\usepackage{soul} % for highlighting text

\usepackage{subcaption}

\usepackage{multirow}
\usepackage{enumitem}

\begin{document}

\title{Reliable and Efficient Data Collection in UAV based IoT Networks}

\author{Joshi Poorvi,~\IEEEmembership{Student Member,~IEEE,}  Alakesh Kalita,~\IEEEmembership{Member,~IEEE,}  Mohan Gurusamy,~\IEEEmembership{Senior Member,~IEEE}

    \thanks{}
    \thanks{}
    \thanks{}
    % \thanks{ Manuscript received January XX, 2021; revised May XX, 2021; January XX, 2021; accepted February XX, 2021. Date of publication February XX, 2021; date of current version July XX, 2021. }
    % \thanks{\emph{Corresponding author: Alakesh~Kalita.}}
    \thanks{J.~Poorvi and  M.~Gurusamy are with the Communications \& Networks Lab, Department of Electrical and Computer Engineering, National University of Singapore. (e-mail: \texttt{e1144005@u.nus.edu}, \texttt{gmohan@nus.edu.sg})}

    \thanks{A.~Kalita is with the ISTD Pillar, Singapore University of Technology and Design, Singapore (e-mail: \texttt{alakesh\_kalita@sutd.edu.sg})}
    % \thanks{}
    % \thanks{}
    % \thanks{}
    % \thanks{}
}
% \markboth{IEEE Internet of Things, ~Vol.~X, No.~X, June~2023}%
% {Hazra \MakeLowercase{\textit{et al.}}: Bare Demo of IEEEtran.cls for IEEE Journals}
\maketitle

\begin{abstract}
 Internet of Things (IoT) involves sensors for monitoring and wireless networks for efficient communication. However, resource-constrained IoT devices and limitations in existing wireless technologies hinder its full potential. Integrating Unmanned Aerial Vehicles (UAVs) into IoT networks can address some challenges by expanding its' coverage, providing security, and bringing computing closer to IoT devices. Nevertheless, effective data collection in UAV-assisted IoT networks is hampered by factors, including dynamic UAV behaviour, environmental variables, connectivity instability, and security considerations. In this survey, we first explore UAV-based IoT networks, focusing on communication and networking aspects. Next, we cover various UAV-based data collection methods their advantages and disadvantages, followed by a discussion on performance metrics for data collection. As this article primarily emphasizes reliable and efficient data collection in UAV-assisted IoT networks, we briefly discuss existing research on data accuracy and consistency, network connectivity, and data security and privacy to provide insights into reliable data collection. Additionally, we discuss efficient data collection strategies in UAV-based IoT networks, covering trajectory and path planning, collision avoidance, sensor network clustering, data aggregation, UAV swarm formations, and artificial intelligence for optimization. We also present two use cases of UAVs as a service for enhancing data collection reliability and efficiency. Finally, we discuss future challenges in data collection for UAV-assisted IoT networks. % and conclude this article.
 
 %we thoroughly investigate present research initiatives aimed at ensuring reliable and efficient data collecting in UAV-assisted IoT networks. These activities are classified according to their goals, which include UAV control, data transfer, energy efficiency, latency, and security. \textcolor{blue}{ Our survey covers critical domain characteristics such efficient data transfer, Wireless Power Transfer (WPT), physical security via Intelligent Reflecting Surfaces (IRS), sensor clustering, and optimization utilizing Reinforcement Learning (RL) and Machine Learning (ML) techniques. We want to deliver useful insights defining the future of UAV-assisted data gathering in IoT networks by presenting an integrated view and highlighting future research areas, supporting seamless and efficient data collection procedures.}

\end{abstract}


%\tableofcontents

\section{Introduction}
\label{sec:introduction}

\IEEEPARstart{I}{nternet} of Things (IoT) is a network of physical objects (also called ``\textit{things}'') that are found in homes, healthcare systems, and industries, to name a few. IoT helps the computational system to interact with the physical world by sensing, communicating, and actuating \cite{ATZORI20102787,6774858}. %In brief, IoT is intended to connect nearly all physical objects, both living and non-living, on Earth's surface \cite{ATZORI20102787,6714496,6774858}. %The nodes or devices in IoT are called \textit{smart} in three aspects: (i) they can \textit{sense or monitor} their surroundings, (ii) they can \textit{communicate} or \textit{exchange} their information with other nodes, (iii) they can \textit{change the behavior} of their surroundings. 
In IoT networks, \textit{sensors} are used for sensing or monitoring the environment, \textit{wireless communication networks} are preferred over wired networks for exchanging information, and  \textit{actuation} is done to change the behaviours of the environment. Thus, IoT can improve the efficiency, safety, and security in different application domains by means of sensing, communication, and actuation. However, due to the usage of resource-constrained devices (aka \textit{nodes}) in terms of processing capacity, storage, and energy in IoT networks and the limited communication range and data speed of the existing wireless technologies,  IoT is not fully used in every immersive application domain and scenarios \cite{6714496}. 

Most of the IoT applications require high-throughput, ultra-reliable, and low-latency wireless communication \cite{bennis2018ultrareliable}. %, alongside ultra-precise and high-resolution wireless sensing capabilities. [\textcolor{red}{cite}]  This presents exciting opportunities for environment and location-aware applications across domains such as smart cities, smart manufacturing, and autonomous driving. 
Nevertheless, traditional terrestrial networks have limitations in providing sensing services over fixed and limited ranges, often obstructed by obstacles that degrade the sensing performance \cite{nguyen20216g}. However, these issues can be resolved with the integration of the \textit{Unmanned Aerial Vehicle} $($UAV$)$ into the communication network as it can not only extend the coverage but also bring the computing resources near to the  IoT devices \cite{aljehani2020uav, 8961914}. 

%Due to the rapid technological growth, devices like mobile user equipment, Internet of Things $($IoT$)$ nodes, etc., have become highly computational intensive. The advanced applications such as online gaming, automatic navigation, video streaming and infrastructure monitoring have imposed a high data traffic and delay-sensitive task requirements on these devices, which have limited battery and computation capabilities. Integrated Wireless Sensor Networks has emerged as potential technology, which requires  high-throughput, ultra-reliable, and low-latency wireless communication, alongside ultra-precise and high-resolution wireless sensing capabilities. This presents exciting opportunities for environment and location-aware applications across domains such as smart cities, smart manufacturing, and autonomous driving. Nevertheless, traditional terrestrial networks have limitations in providing sensing services over fixed and limited ranges, often obstructed by obstacles that degrade the sensing performance. However, these issues can be resolved with the integration of the unmanned aerial vehicle $($UAV$)$ into communication network as it can not only extend the coverage but also bring the computing resources near to the ground devices. 

%\subsection{Background}

IoT devices can transmit data to the \textit{Cloud} via a direct link or by using UAVs as data collectors. Choosing a direct link consumes a lot of energy for IoT devices and necessitates more transmission power to reach a potentially distant Cloud. As a result, their batteries are rapidly depleted. Using UAV data collectors, on the other hand, proved to be a more efficient strategy, allowing for faster and more reliable data collection \cite{8894454}. For example, UAVs support situational awareness and strategic decision-making in military environments by enabling real-time data communication between command centres and the battlefield. Drones\footnote{We use UAV and drones interchangeably in this article.} are used in surveillance operations to send high-resolution imagery and sensor data, boosting monitoring and security capabilities. UAVs can be useful for disaster relief operations because they can quickly access remote disaster areas and gather vital information for effective resource allocation, and with their quick and adaptable data-driven delivery services, UAVs are ready to transform the delivery sector \cite{daud2022applications}.
%With their quick and adaptable data-driven delivery services, UAVs are ready to transform the delivery sector. By autonomously receiving and sending data from distant IoT devices, they also support the emerging Machine Type Communication $($MTC$)$ paradigm.
 Additionally, UAVs with \textit{Edge Computing} $($EC$)$ capabilities can also improve data processing effectiveness. In order to lessen server load and enable real-time decision-making in applications like autonomous navigation, UAVs can pre-process data before transmission.
 %The combination of dependable data collecting and cutting-edge UAV-based data transfer promotes industries while also highlighting the crucial role that data plays in determining the direction of our increasingly technological future.
Furthermore, effective and trustworthy data collecting is the cornerstone of well-informed decision-making and efficient operations across a range of application domains. At the same time, the use of UAVs for data transmission brings in a new era of rapid and accurate data dissemination. %Despite the potential features, there are significant obstacles while using UAVs for next-generation wireless networks. Energy efficiency, security, and reliability are among the primary issues in designing efficient UAV communications. 

%Due to limited onboard energy reserves, energy-aware UAV deployments and operations necessitate sophisticated energy usage and replenishment procedures. Although the UAV-assisted network concept has proven its efficiency in significantly improving the Data transmission process, but the transmission of data towards UAVs is prone to interception by potential malicious eavesdroppers owing to wireless channel's broadcast nature. This security breach poses a risk to data security and confidentiality. Additionally, the data transfer process is subject to various types of risks, such as spoofing, and unauthorized changes in the storage of acquired data, which may raise concerns about data integrity. As a result, several security measures, such as cryptographic approaches or a physical layer security approach, are necessary to assure a particular level of data transfer reliability. Furthermore, despite their limited resources, UAVs assist IoT devices in overcoming mobility issues and navigating complex deployment locations \cite{li2022taskpoi}.

\subsection{Motivation}

Reliable and Efficient \textit{Data Collection} $($DC$)$ in IoT is critical because it improves operational efficiency by automating sensor data collecting, %removing the need for human collection. It gives 
provides accurate real-time insights, %allowing organizations to monitor and fix issues quickly. %The data obtained aids in predictive maintenance and quick decision-making by tracking the operation of equipment connected to the IoT network. This not only 
improves the productivity, scalability and resiliency of the underlying systems.  % and longevity of the devices, but it also monitors ambient parameters such as humidity, temperature, movement, and air quality, providing a foundation for tracking physical work conditions to minimize disasters.
%Although the importance of UAVs is expanding, their deployment as
However, efficient and reliable data collection is challenging in UAV-assisted IoT networks. %due to a variety of challenges.
Because of the dynamic nature of UAVs, environmental conditions, and the distance between the UAV and IoT devices, unstable connections could arise. Interference, signal fading, and noise can all create bit mistakes during data transmission, compromising data quality. The dynamic topology of the UAV network, as well as the demand for low latency services and consideration of Age of Information $($AoI$)$, hinder routing optimization. Furthermore, security threats such as data manipulation and theft, especially when malevolent organizations are present in the network, offer substantial hurdles. As a result, addressing these difficulties is critical for effective and dependable data collecting in these networks.

UAV installations introduce increased interferences, especially when IoT devices broadcast data over identical communication bands, potentially leading to communication failures, increased battery consumption for retransmission, and extended DC durations. As a result, rigorous interference consideration in channel modelling is critical \cite{mkiramweni2018game}\cite{yang2021sum}. Another significant barrier between UAVs and IoT is energy constraints. UAVs rely on rechargeable batteries, severely limiting mission duration due to limited battery capacity \cite{abeywickrama2018comprehensive}. Mobility, communication capability, and the energy consumed by their integrated circuits significantly affect the energy consumption of UAVs. In addition, IoT devices, which are similarly constrained by energy and not possible to replace their batteries frequently, rely on periodic data transmission to UAVs and erratic movement, which exacerbates the difficulty \cite{zhong2020deep}. %Overcoming this barrier is critical for prolonging the DC process's operation period, demanding sophisticated charging technologies or techniques to sustain continuous functionality \cite{zhong2020deep}. 
Due to prevalent Non-Line-of-Sight (NLoS) conditions between UAVs and IoT devices, the heterogeneous character of these networks creates complex channel modelling, making Air-to-Ground (A2G) channels prone to blocking \cite{huang2021empirical}. To precisely simulate such channels, many factors such as inclination angle, UAV altitude, transmission environment, obstacles, and interferences must be considered, requiring extensive simulations and measurements for validation.

The necessity for constant UAV movement makes efficient DC even more difficult \cite{gul2022uav}. To enable successful data relay and collection, UAVs must constantly traverse the neighbourhood of terrestrial IoT devices, maintaining persistent contact with surrounding UAVs. However, designing ideal UAV trajectories is difficult due to various constraints such as UAV energy consumption, existing impediments, IoT device movement, and the mobility of other UAVs. Dynamic trajectory control techniques for UAVs are becoming increasingly important to ensure a long connection lifespan while sufficiently covering the target region of interest \cite{chen2020toward}. As the UAV-assisted terrestrial network idea improves the DC process, it remains vulnerable to potential data interception during transmission to UAVs due to the broadcast nature of wireless channels \cite{huang2021energy}. This security flaw jeopardizes data security and confidentiality, emphasizing the need for robust security protocols to provide protection against potential threats such as man-in-the-middle attacks and unauthorized data alterations \cite{basan2021self}. To summarize, tackling these complex difficulties together is critical to enabling seamless and efficient and reliable data collecting in UAV-assisted IoT networks, which necessitates novel solutions and a complete strategy. 

To overcome these challenges for DC in UAV-assisted IoT networks, researchers proposed different schemes considering different objectives such as UAV control, data transfer, energy efficiency, latency efficiency, and security. Reliable and efficient data collection from IoT networks helps in achieving high reliability, scalability, sustainability, and resiliency in different IoT application domains. Numerous studies in the literature have delved into the extensive research dedicated to this direction. However, a clear roadmap for achieving reliable and efficient data collection in UAV-assisted IoT networks remains elusive. Therefore, a comprehensive analysis of the state-of-the-art methods from different domains can bring valuable insights to shape the future of UAV-assisted data collection in IoT networks and helps in answering the questions like ``\textit{\textbf{How to design advanced data collection techniques in UAV-assisted IoT networks?}}'' Therefore, in this survey article, we extensively study the existing works towards reliable and efficient data collection in UAV-assisted IoT networks.% considering different aspects such as  communication protocols, security, UAV control and \textcolor{red}{need to add more} 

\subsection{Comparison with Existing Surveys}

Many research groups across the world are currently working on energy-efficient and delay-sensitive data collection and transmission. The work in \cite{8660516} summarises analytical frameworks that are expected to play an important role in designing future UAV-based wireless networks that enable network operators to leverage UAVs for various application scenarios. They have considered different classes of UAVs and their roles in different applications. The paper \cite{8411465} reviews measurement techniques for UAVs as low-altitude platforms. Established analytical UAV channel models and stated important issues related to airframe shadowing, non-stationary channels, and diversity techniques in UAV communications. \cite{9044378} provide a comprehensive survey of routing protocols for aerial UAV networks together with communication and application architectures, mobility models, and related analysis on routing protocol design. \cite{jawhar2017communication} discussed some energy-efficient UAV-based Data Collection via Wireless Sensor Networks (WSNs), as well as numerous networking layer protocols utilized by UAV-based systems. Regarding UAV energy consumption, interference, and spectral efficiency, \cite{sekander2018multi} assessed the advantages of multitier UAV network architecture over conventional single-tier UAV networks. The focus of \cite{8843851} was on maximising energy efficiency in UAV networks and how UAVs are employed in the localization and recognition of UAV nodes. The \cite{hayat2016survey} investigates the characteristics and demands of UAV communication systems for urban field applications from a networking and communication perspective. Specific criteria examined by the authors included the minimum quantity of data to be transmitted across the network, mission-relevant network characteristics, and quality of service requirements. In addition, common networking requirements, including scalability, security, privacy, safety, adaptability, and connectivity, are highlighted. Bithas et al. \cite{bithas2019survey} performed a comprehensive analysis of machine learning (ML) methods used to improve UAV-based communications. Fu et al. \cite{fu2022toward} investigated deep learning algorithms for addressing the issue of low energy efficiency in wireless UAV networks.

In terms of the integration of security factors into UAV-assisted networks, \cite{8675384} have explored a wide range of UAV-assisted network and cyber-physical security challenges. Furthermore, \cite{nguyen2021uav} thoroughly examined UAV-based DC schemes in WSNs while taking into account various security challenges associated with such systems. The emphasis was on incorporating security and Wireless Power Transfer (WPT) ideas into the UAV-based DC process. Because ground nodes and UAVs have limited energy capacity, \cite{pakrooh2021survey} a survey on services based on UAV-assisted IoT environments. The many techniques addressed are classified into multiple categories, including UAV-assisted power transfer, UAV-assisted Mobile Edge Computing (MEC), and UAV-assisted communications. \cite{alzahrani2020uav} investigated the benefits of the UAV aid paradigm, including the provision of an aerial energy source to ground network users, the provision of flying monitoring support, and the establishment of a communication bridge between isolated networks. In \cite{MESSAOUDI2023103670} an extensive survey explored current research on optimal UAV-based data collection schemes, categorizing them into multiple domains. %, including Movement Control, Data Transfer, Energy Efficiency, RIS, WPT and more. This structured classification provides valuable insights into the UAV-based data collection landscape.
In \cite{9779853}, a comprehensive UAV-assisted data collection solution was presented, addressing critical technologies like joint sensing, efficient sensor access, and optimal routing in WSNs.% Future research directions highlighted the potential of efficient multiple access and integrated joint sensing for advancing UAV-enabled data collection methodologies. 

In contrast to existing survey works on UAV-aided data collection, our survey takes a comprehensive approach by integrating key components critical to this domain. While existing surveys often focus on specific aspects, we acknowledge the need to consider a holistic spectrum for a more robust understanding. Specifically, this survey work differs from the existing surveys as follows.

\begin{itemize}
   % \item Reliable and Efficient data collection together in UAV-assisted IoT networks. 
   % \item Reliable and Efficient data collection in IoT device-to-IoT device, IoT device-to-UAV, UAV-to-UAV, and UAV-to-Base Station communication scenarios.
    \item This work emphasizes reliable and efficient data transfer mechanisms together as a fundamental requirement in UAV-aided data collection from IoT networks. 
    \item We cover optimizing data transmission protocols that improve packet delivery latency and network throughput, secure data transmission schemes, sensor clustering and data aggregation techniques, and optimized path planning and coverage schemes as the key aspects for enhancing overall data collection efficiency and reliability in UAV-assisted IoT networks. We have categorized these existing schemes to provide a deeper understanding of reliable and efficient data collection.
    \item We consider IoT network, UAV network, and UAV-IoT network for ensuring reliable and efficient data collection across various communication scenarios: IoT device--to--IoT device, IoT device--to--UAV, UAV--to--UAV, and UAV--to--Base Station.
    % \item Sensor clustering, a vital technique in data collection, is also highlighted. Our survey delves into intelligent sensor grouping methodologies, optimizing data relevance, power consumption, and collection efficiency.
     %\item We also underscore the importance of Wireless Power Transfer (WPT) mechanisms, offering insights into sustaining UAV operations through wireless energy replenishment. This is vital in extending UAV operational capabilities and data collection durations, a critical consideration often overlooked in existing surveys.
    % \item Intelligent Reflecting Surfaces (IRS) aided security, is another key focus area. By intelligently redirecting signals and enhancing communication robustness, IRS plays a significant role in ensuring secure data transmission, a facet essential for modern data collection frameworks.
    % \item In the realm of optimization, we integrate cutting-edge Reinforcement Learning (RL) and Machine Learning (ML) algorithms. These algorithms drive optimizations in UAV path planning, resource allocation, and data collection strategies, aiming for efficiency and efficacy.
\end{itemize}

%In summary, our survey amalgamates these critical aspects: efficient data transfer, Wireless Power Transfer (WPT), physical security via IRS, sensor clustering, and optimization through RL and ML algorithms. By presenting an integrated view, we strive to provide a thorough understanding and pave the way for advancements in UAV-aided data collection. 
\color{black}

%\textcolor{red}{Need to add a comparison based Table Ref[13-27]}

% \subsection{Scope and Objective}
% \textcolor{red}{sir what i have to include in this?}

\color{black}

\subsection{Contribution and Organization}
In this article, we comprehensively review the existing works in UAV-assisted IoT networks, focusing on reliability and efficiency. Reliable and efficient data collection is necessary to improve the scalability, reliability, and resiliency in UAV-assisted IoT environments. To achieve this, at the beginning, we briefly discuss how UAV and IoT can be integrated for DC and what are the data collection methods with their pros and cons. After that, we first categorised the existing works based on their primary objectives--namely, reliability and efficiency. Subsequently, we further sub-categorized the works into different categories based on their core contributions towards different facets such as communication protocols, security, UAV control, etc. These concise categorisations of the existing articles will help the research communities to understand and design advanced data collection techniques for reliable and efficient data collection in UAV-assisted IoT networks. Furthermore, we discuss a few applications and use cases of UAV-assisted IoT networks, considering UAVs as a service. We also discuss future research directions and open issues to UAV-based reliable and efficient data collection.  The main contributions of this article are summarized as follows.
\begin{itemize}
    \item At the beginning, we briefly discuss UAV-based IoT networks, considering communication and networking aspects. We also provide a detailed discussion of different data collection methods, their advantages and disadvantages, and performance metrics for reliable and efficient data collection in UAV-aided IoT networks.
    \item We review existing work related to reliable data collection in UAV-assisted IoT networks, categorizing them into (i) Data Accuracy and Consistency, (ii) Network Connectivity, and (iii) Data Security and Privacy. Furthermore, we further subdivide these categories to provide a more in-depth understanding of reliable data collection.
     \item We briefly discussed the existing works on efficient data collection in UAV-assisted IoT networks, categorizing them into (i) Trajectory and Path Planning,  (ii) Sensor network clustering, (iii) Data aggregation,  (iv) UAV Swarm Formations and (v) AI Optimization techniques. We further subdivide these categories to provide a more in-depth understanding of efficient data collection.
    % \item  We highlight key approaches like Genetic Algorithms (GAs), Particle Swarm Optimization (PSO), Simulated Annealing (SA), and Ant Colony Optimization (ACO). These algorithms find wide-ranging applications in machine learning, function optimization, logistics, and more, offering efficient problem-solving solutions.
     \item Additionally, we discuss two use cases of UAVs targeting to improve the reliability and efficiency of data collection, considering UAVs as a service. Finally, we provide a brief discussion on future challenges related to UAV-based reliable and efficient data collection and conclude this article. %Moreover, possible solutions are provided along with recommended references.
     
\end{itemize}

\begin{figure}[!t]
    \centering 
    \includegraphics[width=\columnwidth]{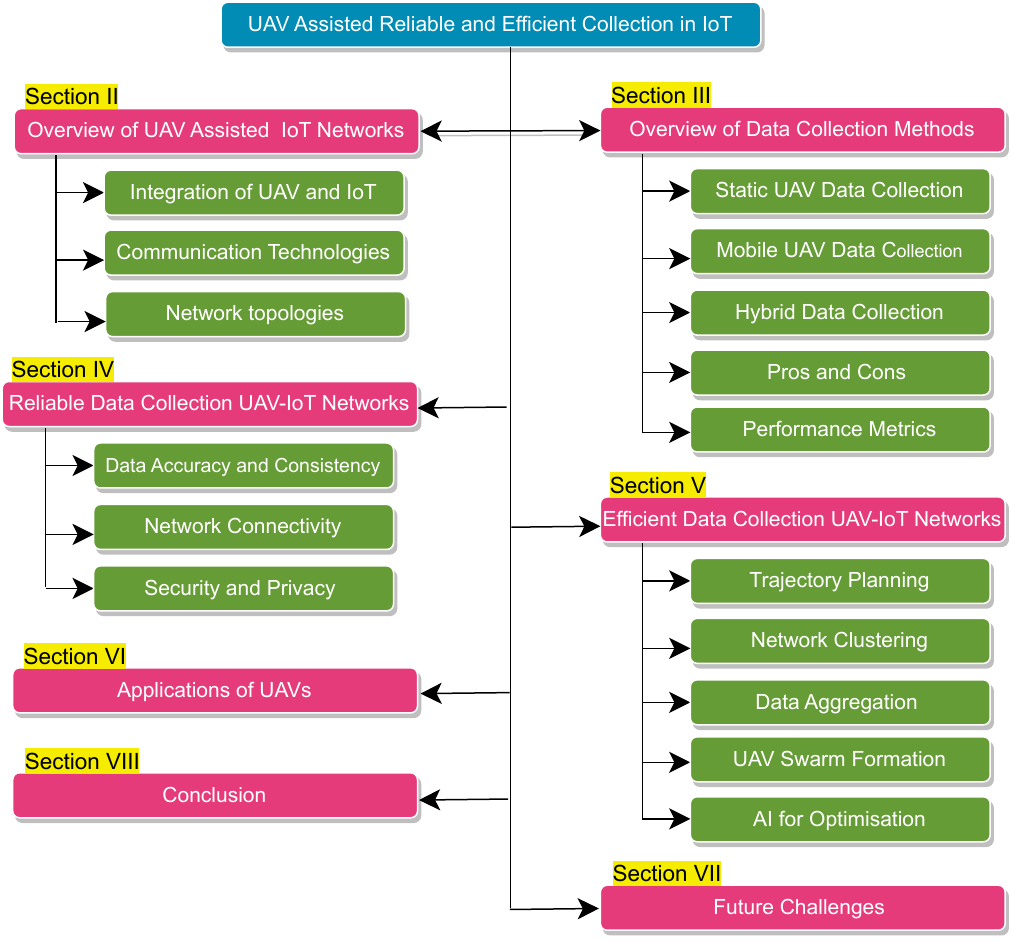}
    \caption{Organization of this article}   
    \label{org}
\end{figure}

As shown in Fig.~\ref{org}, the remainder of this article is organized as follows: Section~\ref{sec:fundamentals} discusses the overview of UAV-assisted IoT networks, and Section~\ref{sec:fundamentals2} provides the overview of UAV's data collection methods. In Section~\ref{sec:reliable}, we discuss the works related to reliable data collection techniques and in Section~\ref{sec:efficient}, we briefly discuss the existing techniques related to efficient data collection in UAV-assisted IoT networks. %We also discuss the existing works that consider both reliable and efficient data collection in UAV-assisted IoT networks in Section~\ref{sec:both}. 
We discuss the other applications of UAVs considering UAVs as a service in Section~\ref{sec:applications}. Next, we briefly discuss the future challenges and their probable solution for reliable and efficient data collection in UAV-assisted IoT networks in Section~\ref{sec:challenges-solutions}. Finally, we conclude this article in Section~\ref{sec:conclusion}.

\section{Overview of UAV-based IoT Networks}
\label{sec:fundamentals}
In this section, we discuss several key aspects of integrating UAVs and IoT for data collection. We begin by explaining the components of UAV-based IoT networks, followed by exploring the various communication technologies that can be used for such UAV-based IoT networks. Additionally, we also discuss various network topologies, addressing the distinct communication needs of IoT nodes, UAVs, and base stations.

% \subsection{Introduction to UAVs}
% \subsection{Basics of IoT}
\subsection{Integration of UAVs and IoT}

Integrating UAVs and IoT for data collection involves combining the capabilities of drones and IoT devices to gather, transmit, and analyze data in a seamless and efficient manner. The network components of UAV-supported data collection encompass several essential elements, including sensors, UAVs, control stations, base stations/access points (BS/AP), and servers, which are discussed as follows:
\begin{enumerate}
    \item \textbf{Sensor Deployment:} Sensors are strategically deployed to monitor and sense the environment, collecting data relevant to specific applications. The type of sensors employed varies based on the application, such as temperature and humidity sensors for forest monitoring. Note that sensors are attached to the IoT devices, and IoT devices can form networks among themselves and also connect to the UAV networks.
    \item \textbf{UAV Data Collection:} UAVs fly over the deployed sensors, effectively collecting the sensed data. The collected data is then either transmitted to the cloud server or handled by themselves for processing and storage.
    \item \textbf{Control Station: } A control station is responsible for planning and coordinating the flight path of UAVs to ensure efficient and comprehensive data collection.
    \item \textbf{Base Station/Access Point:} BS/AP acts as a hub, directly collecting sensing data from surrounding sensors and UAVs and aggregating the data for further processing.
    \item \textbf{Data Processing Server:} Clouds store and process the sensor data, facilitating subsequent analysis and insights.
\end{enumerate}

\begin{figure*}[!t]
    \centering 
    \includegraphics[width=.7\linewidth]{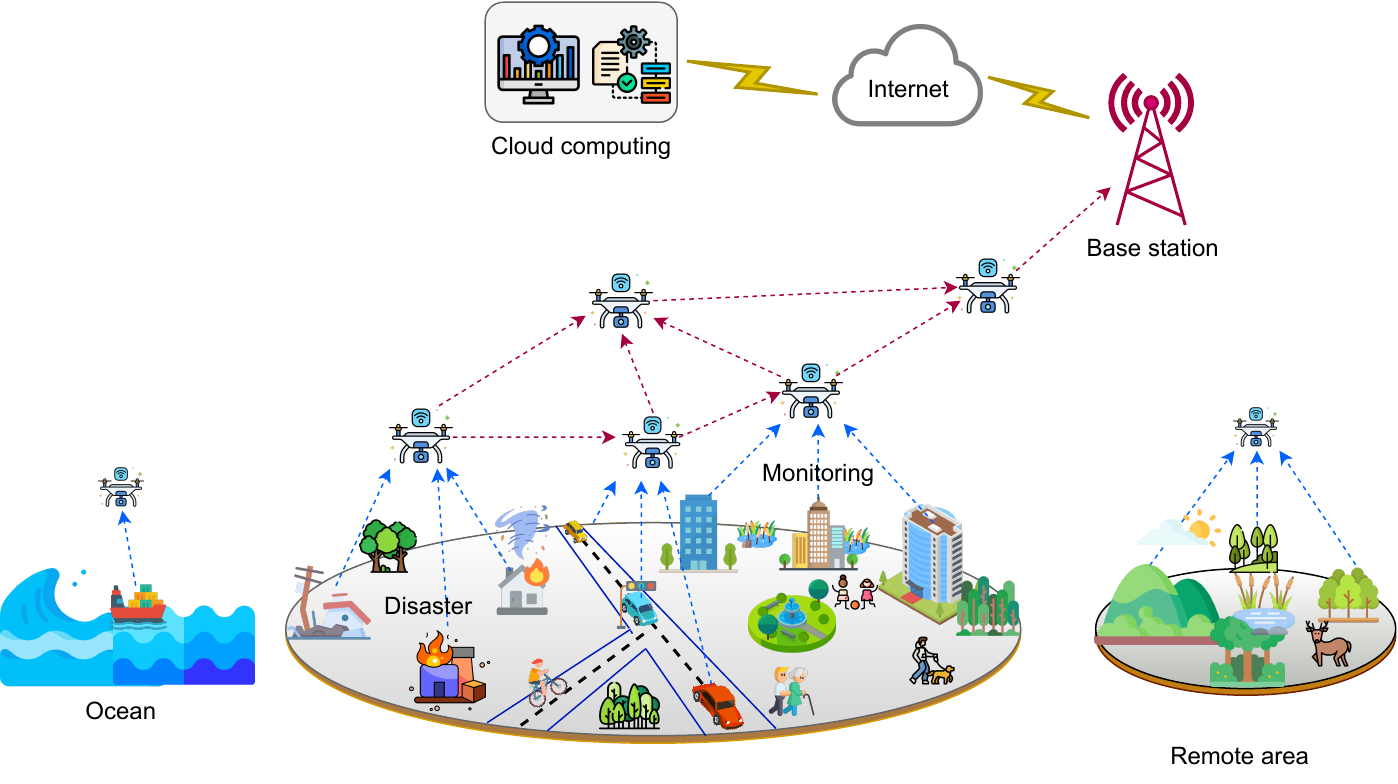}
    \caption{Applications of UAV-assisted IoT Networks}    
    \label{fig_1:Network}
\end{figure*}

As in Fig.~\ref{fig_1:Network} the UAV-assisted data collection scenario, first the UAV is launched from a control station and navigates over the sensors, gathering vital data such as potential hydrogen levels, salinity in the ocean, wildlife and fire status in forests, or traffic and weather information in smart cities. The collected data is transmitted to the BS/AP and then returned to the control station, aiming for efficiency, energy conservation, and minimal delay during the data collection. In this scenario, the need for reliable and efficient data collection is essential for decision-making accuracy, promptness in responding to rapidly changing conditions, energy conservation to maximize the UAV's operational time, safety in areas of risk, and scalability to cover multiple locations at the same time.

Furthermore, standard communication technologies like Wireless Fidelity (WiFi) and Long Term Evolution $($LTE$)$ are integrated into UAV-assisted data collection. Some communication technologies can serve as a gateway, extending coverage to rural and remote areas, while some other can support extensive coverage and operates with low power and cost-effectiveness, making them suitable for applications such as soil detection, city open water monitoring, and air quality assessment in UAV-assisted data collection. However, the selection of communication technologies and protocols greatly impacts the overall performance of the UAV-assisted IoT networks and deployment cost. Therefore, in the next section, we discuss the different wireless communication technologies that can be used for IoT device--to--IoT device, IoT device--to--UAV, UAV--to--UAV, and UAV--to--BS communication scenarios.
\begin{table}[!ht]
    \centering\footnotesize  \caption{Nomenclature} \label{abbr} \rowcolors{2}{gray!50!blue!20}{gray!20!blue!1}
    \begin{tabular}{|l|p{6.0cm}|}
    \toprule
    \rowcolor{lightgray} \textbf{Acronym} & \textbf{Abbreviation}\\
    \midrule

6TiSCH & IPv6 over Time Slotted CHannel\\
        A2G & Air to Ground\\
        ACK & Acknowledgement \\
        AI & Artificial Intelligence\\
        AoI & Age of Information\\
        BR & Border Router\\
        BS & Base Station\\
        CH & Cluster Head\\
        CN & Centralised Nodes\\
        CSI & Channel State Information\\
        DC & Data Collection\\
        DRL & Deep Reinforcement Learning\\
        EC & Edge Computing\\
        ED & End Devices\\
        EE & Energy Efficiency\\
        ET & Energy Transmitter\\
        GD & Ground Devices\\
        IIoT & Industrial Internet of Things\\
        IoT & Internet of Things\\
        LoS & Line of Sight\\
        MAC & Medium Access Control\\
        MEC & Mobile Edge Computing\\
        MA & Mobile Agent\\
        MANET & Mobile Ad-Hoc Networks\\
        MPR & MultiPoint Relay\\
        NB-IoT & Narrow Band Internet of Things\\
        NLoS & Non line of Sight\\
        NOMA & Non-Orthogonal Multiple Access\\
        PLS & Physical Layer Security\\
        QoS & Quality of Service\\
        TDMA & Time Division Multiple Access\\
        UAV & Unmanned Aerial Vehicles\\
        UGV & Unmanned Ground Vehicle\\
        VANET & Mobile Ad-Hoc Networks\\
        WPT & Wireless Power Transfer\\
        WSN & Wireless Sensor Network \\

\bottomrule
\end{tabular}
   
\end{table}
\subsection{Communication in UAV-assisted IoT Network}

%Various communication protocols facilitate data transfer among drones, satellite connections, and aerial control units via Air-to-Air (A2A) communications, as well as between airborne and ground-based systems through Air-to-Ground (A2G) methods. Ground-based stations encompass independent control units, larger servers, Internet gateways, and edge computing devices. The networking aspect, can be examined from multiple aspects in terms of communication, computation, and scheduling needs and limitations. Many aerial monitoring platforms traditionally rely on ground-based or web-based servers for bulk data processing. In this setup, drones gather and transmit raw data to processing units. However, emerging approaches, such as on-the-fly processing employing lightweight embedded GPUs/TPUs, Mobile Edge Computing (MEC), and fog computing, are gaining traction in recent years. These methods aim to accelerate and achieve near-real-time data processing \cite{shi2016edge}\cite{dastjerdi2016fog}.

Communication protocols facilitate data transfer among IoT devices, UAVs, and UAV-IoT networks.  In many scenarios, the operational area of an IoT network can be extended far beyond the Line of Sight (LoS) communication range by using UAVs. Thus, employing a network of interconnected UAV platforms is imperative to ensure seamless connectivity.  A critical design consideration revolves around selecting the most suitable wireless technology (e.g., WiFi, LTE bands) that can offer adequate capacity and meet acceptable Quality of Service (QoS) standards. This choice is evaluated based on both technical feasibility and economic viability. %Efforts have been dedicated to establish a nationwide high-speed broadband wireless network, especially focused on public safety communications. For instance, FirstNet represents a notable initiative, aiming to deploy, operate, maintain, and enhance the first high-speed, nationwide wireless broadband network dedicated to public safety. This model holds promise and could be extended to UAV networks in the near future \cite{6599064}. The potential broadband wireless technologies encompass WiFi, 4G Long Term Evolution (LTE), 5G (aligned with the 3rd Generation Partnership Project (3GPP) standard on 5G communication for drones). %, satellite communications, and dedicated public safety systems like TETRA and APCO25.
Additionally, technologies like Long Range Wide Area Network (LoRaWAN) enable long-range, low-power communications, especially suitable when high throughput is not the primary concern \cite{delafontaine2020drone, saraereh2020performance}. We discuss these types of most widely used communication technologies as follows.

\subsubsection{Bluetooth Low Energy $($BLE$)$}
BLE is a wireless technology used in medical, environmental, safety, and energy sectors, contributing to technological improvements \cite{6411303}. Unlike traditional Bluetooth, BLE uses minimal power to ensure cost-effectiveness while preserving communication range. Furthermore, BLE prioritizes privacy and flexibility when it comes to openly sharing data. Its effectiveness is particularly noticeable when delivering tiny amounts of data, such as temperature sensor readings, GPS locations, and acceleration details, making it ideal for agricultural applications. Regarding location detection, BLE outperforms Wi-Fi, which fails to penetrate solid objects adequately. Furthermore, BLE provides highly accurate proximity detection within a range of 30 meters, which can be extended up to $100$ meters with extra antennas.  Ultimately, BLE's superior advantage lies in its lower power consumption and cost-effectiveness compared to Wi-Fi \cite{prasad2018ble}.

\subsubsection{6TiSCH}
The 6TiSCH (IPv6 over Time Slotted Channel Hopping (TSCH) mode of IEEE 802.15.4e) wireless protocol stack is released to offer high-throughput, low and bounded latency, energy efficient, and reliable communication in mission-critical IoT applications. 6TiSCH networks are built on the top of IEEE 802.15.4e standard \cite{7460875}, which is a low-power wireless communication standard that introduced TSCH MAC mode to provide stringent requirements such as high reliability, higher throughput, delay-bounded, and energy-efficient communication in resource-constrained node based multi-hop IoT networks \cite{8278196}. 6TiSCH can be used to enable real-time monitoring, such as assessing temperature, humidity, and other environmental factors, where sensors are placed all over anywhere on the Earth's surface. The sensors attached to IoT nodes deliver real-time data to the root node (aka Border Router (BR)) using the Routing Protocol for Low power and Lossy network (RPL) \cite{RFC-RPL} of the 6TiSCH protocol stack. The BR can process such data by itself or transmit it to the \textit{Cloud} using the Internet for further processing. %6TiSCH provides....

\subsubsection{Wi-Fi}
The primary communication technology utilized in the communications of many commercial UAVs is Wi-Fi (IEEE 802.11 series), especially when connecting to a ground station for tasks like command and control commands in the uplink and video streaming in the downlink \cite{8612108}.  Wi-Fi is chosen for its cost-effectiveness, scalability, and affordability. Additionally, Wi-Fi-based UAV networks have the capability to serve as wireless backhauling systems \cite{7470932}. In the context of inter-drone communication, Wi-Fi can power the connections, provided that one node is designated as AP to establish a local WLAN. However, this AP may or may not offer Internet access. One limitation of Wi-Fi is its handling of mobility and hand-off between base stations, constraining the operational range of drones to a few miles within direct access to the AP. Despite Wi-Fi's throughput being theoretically lower than LTE and 5G -- ranging from 54 Mbps for 802.11a to as high as 2.4 Gbps for 802.11ax, it typically suffices for a wide array of applications, including real-time high-resolution video streaming. %However, in scenarios demanding long-range connectivity, Wi-Fi is overshadowed by licensed wireless systems whenever such networks are available. Some drones adopt custom communication protocols built upon the Wi-Fi infrastructure. For instance, the XFold spy x8 KDE U3 drone by Xfold Rigs \cite{} is equipped with a Futaba Commercial 14-channel, demonstrating this approach.
%Communication between sensors and cluster heads, interactions among UAVs, and communication between UAVs and ground stations are all used in Unmanned Aerial Wireless Sensor Networks (UAWSNs). Maintaining survivability and reliability in UAWSNs requires multiple communications to offset potential packet loss caused by link failures between UAVs and sensor nodes. 
IEEE 802.11 Medium Access Control (MAC) protocol is commonly utilized for communication between UAVs and ground stations due to its appropriateness for long-distance communication, high bandwidth handling, and fast data rates. Conversely, inter-UAV communications present difficulties due to fluctuating link quality induced by UAV mobility, dynamic topological changes, and varying communication distances between the communicating nodes. %To the aforementioned communication links, many communication protocols are used. \color{red} Table \color{black} summarizes and compares various communication methods utilized in UAWSNs for various purposes, taking physical characteristics and communication parameters at the physical and data link levels into account.

\begin{table*}[!t]
\centering
\caption{\footnotesize{Comparison of wireless communication technologies}}
\begin{flushleft}
\footnotesize{\textquotedblleft $O_1$\textquotedblright \; signifies \textit{Reliability}. \textquotedblleft $O_2$\textquotedblright \; signifies \textit{Security}. \textquotedblleft $O_3$\textquotedblright \; signifies \textit{QoS}. \textquotedblleft $O_4$\textquotedblright \; signifies \textit{Interoperability}.}
\end{flushleft}

\label{protocol}
\resizebox{.8\textwidth}{!}{%

\begin{tabular}{cccccccccc}

\hline \hline
\rowcolor[HTML]{8FEDDD} 
 \textbf{Technology} & {\color[HTML]{000000} \textbf{Standard}}  & \textbf{Topology} & \textbf{Speed (kbps)} & \textbf{Range (meters)} & {\color[HTML]{000000} \textbf{$O_1$}} & {\color[HTML]{000000} \textbf{$O_2$}}  & {\color[HTML]{000000} \textbf{$O_3$}} & {\color[HTML]{000000} \textbf{$O_4$}}  \\ \hline \hline

Zigbee* & IEEE 802.15.4  & Mesh & 250 & 1000 & \ding{55} & \checkmark & \ding{55} & \checkmark  \\ \hline

\rowcolor[HTML]{EFEFEF} 
6TiSCH & IEEE 802.15.4e  & Mesh & 250 & 1000 & \checkmark & \checkmark & \checkmark & \checkmark \\ \hline

RFID & ISO/IEC 18000 & Point-to-Point & 1 & 12-100 & \checkmark & \checkmark & $-$ & \checkmark  \\ \hline

\rowcolor[HTML]{EFEFEF} 
NFC & ISO/IEC 18092  & Point-to-Point & 0.424 & 0.02 & \checkmark & \checkmark    & $-$ & \checkmark  \\ \hline

SigFox* & SigFox  & Star & 0.1 & 10000 & \checkmark & \checkmark   & $-$ & \checkmark  \\ \hline

\rowcolor[HTML]{EFEFEF} 
NB-IoT* & 3GPP & Star & 60 & 10000 & \checkmark & \checkmark & $-$ & \checkmark  \\ \hline

WiFi & IEEE 802.11  & Point-to-Point & 20000 & 100 & \ding{55} & \checkmark & \ding{55} & \checkmark  \\ \hline
\rowcolor[HTML]{EFEFEF} 
LoRaWAN & LoRaWAN TS1-1.0. 4  & Point-to-Point  & 0.3 & 10000 & \checkmark & \checkmark & \checkmark & \checkmark  \\ \hline

BLE & IEEE 802.15.1  & Point-to-Point & 1024 & 100 & \checkmark & \checkmark & \ding{55} & \checkmark  \\ \hline \hline
\end{tabular}%
}
\begin{flushleft}
\footnotesize{\textquotedblleft ~$*$~\textquotedblright \;\; signifies that the protocol is proprietary;} \footnotesize{\textquotedblleft ~$-$~\textquotedblright \;\; signifies that the features not available;} \footnotesize{\textquotedblleft \checkmark~ \textquotedblright \;\; signifies that the features are fully/partially available;} \footnotesize{\textquotedblleft \ding{55}~ \textquotedblright \;\; signifies that the features are not availabl.}
\end{flushleft}
\end{table*}

\subsubsection{LTE}

LTE technology improves airborne connectivity for drones by extending communication beyond Line of Sight (LoS), increasing throughput, and strengthening network connectivity through robust hard and soft hand-over procedures \cite{lin2018sky}. There has been a flurry of activity in recent years focused on exploiting terrestrial LTE networks to give UAV connectivity. In January 2017, the FAA and NASA launched a joint initiative in the United States to develop an LTE-based system for UAVs \cite{lin2018sky}. In March 2017, the 3GPP organized a study group to investigate expanded LTE support for aerial vehicles to understand the potential of  LTE's on tiny UAVs \cite{zeng2018cellular}. However, one significant disadvantage of using LTE and comparable cellular technologies is the requirement to register drone transmitters with a service provider. The licensing requirement raises operational expenses and limits drone operations to areas served by the individual service provider, which is why cellular systems are less popular for drones than Wi-Fi. Furthermore, LTE propagation planning is predominantly geared toward ground users, resulting in inferior propagation maps for aerial nodes. This disparity necessitates considerable adjustments in LTE radio planning to properly service UAV networks, particularly as they expand into huge networks at higher altitudes with varied topologies.

\begin{figure}[!t]
    \centering 
    \includegraphics[width=.6\columnwidth]{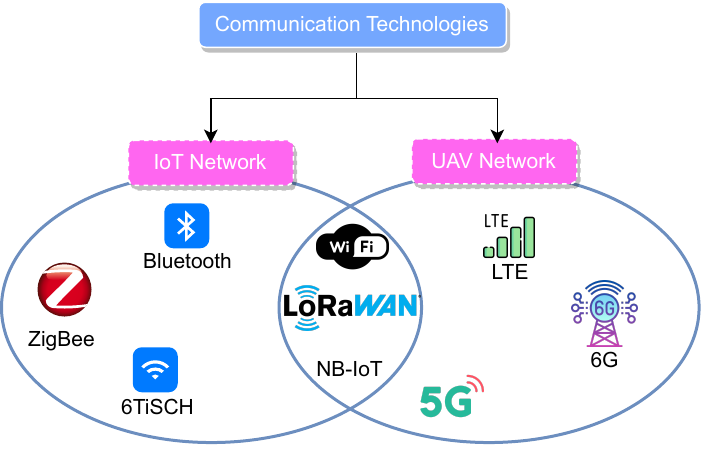}
    \caption{Communication standards for UAV and IoT Networks}    
    \label{fig:Comm}
\end{figure}

\subsubsection{Fifth Generation (5G)}

The 5G network is being investigated for drone communications, particularly when larger bit rates are necessary \cite{li2018uav}. It presents the IoT concept for drones, imagining a scenario in which a drone functions as a ``thing''. UAV-assisted communications provide several prospective benefits, such as on-demand deployment, great flexibility in network reconfiguration, and the construction of LoS communication links. Drones can operate as 5G radio stations in certain scenarios, extending the coverage of 5G networks for ground users, particularly for important applications such as public safety and disaster management \cite{naqvi2018drone,sekander2018multi}. 

\subsubsection{Sixth Generation (6G)}

The need for increased data speed and handling a larger volume of devices is ever-growing, driving the development of 6G to cater to these demands. Positioned as the upcoming wireless communication generation, 6G is anticipated to offer intelligent, secure, dependable, and boundless connectivity at rates 100 times faster than 5G \cite{8869705}. Similar to the support 5G provides, we anticipate that 6G networks will better fulfil the varied requirements of UAV networks, encompassing low latency, reliability, and energy efficiency, making aerial nodes an essential component of 6G networks. Additionally, 6G is envisioned to emphasize network intelligence, a pivotal feature that can significantly aid various connectivity-related applications \cite{mozaffari2021toward}. For instance, it could be critical in a blockchain-based solution designed for UAV communications \cite{aggarwal2020blockchain}. Anticipated challenges posed by the futuristic notion of a connected sky include issues like high mobility, interference, and connection to down-tilted antennas. Integrating aerial nodes with terrestrial nodes is projected to be instrumental in addressing these challenges and elevating the overall 6G user experience.

\subsubsection{LoRaWAN}

It is a wireless communication protocol designed for IoT applications. The network architecture employs a star-of-stars topology, with gateways relaying messages between end devices and a central network server. These gateways connect to the network server via regular IP connections and operate as a transparent bridge, transforming RF packets to IP packets and vice versa\cite{8019271}.  LoRaWAN allows for bidirectional communication, end-to-end security, mobility, and localisation. It features three separate end-point device classes to satisfy the various needs reflected in the diverse range of applications. LoRaWAN can be especially useful in the context of UAV-assisted systems due to its power efficiency, making it a feasible option for drones that normally run on batteries. The UAV can collect data from numerous sensors and send it over the LoRa network. While the raw LoRa protocol can be used for UAV-to-UAV communications, LoRaWAN might be utilized for UAV-to-ground communications to take advantage of pre-existing LoRaWAN networks. It is crucial to remember, however, that LoRaWAN may not be ideal for real-time or control-heavy applications\cite{9530545}.

\subsubsection{NB-IoT}
Narrowband IoT (NB-IoT) is a specialized cellular network technology that has been fine-tuned to meet the rising demands of the IoT landscape. It optimizes spectrum consumption by using a subset of the LTE standard.%, OFDM modulation for downlink communication, and SC-FDMA for uplink communication within a narrow 200kHz bandwidth. 
The deliberate selection of a restricted bandwidth is a strategic decision to improve spectrum efficiency and permit efficient data transfer, which is critical in IoT applications. NB-IoT addresses significant issues in IoT deployments by focusing on indoor coverage, cost-efficiency, and battery life extension. It achieves an excellent balance by considerably improving power usage in user devices, hence increasing system capacity, particularly in difficult-to-cover areas. %Furthermore, as compared to existing GSM/GPRS technologies, NB-IoT technology simplifies system architecture while staying cost-effective, with estimated initial module costs equivalent to GSM/GPRS. As demand for NB-IoT develops, its underlying technology is likely to become more affordable, accelerating its use in IoT applications. 
Because of its power economy, NB-IoT has great potential in the context of UAV-assisted systems, making it an ideal alternative for drones that normally function on battery power. Drones with NB-IoT capabilities can easily collect data from a wide range of sensors and send it over the network, ensuring continuous and dependable connectivity. Despite the fact that alternatives such as LoRaWAN are recognized for their low power consumption and large coverage, the advantages of NB-IoT in terms of low latency, strong security features, and high interoperability make it a tempting choice, albeit at a somewhat higher cost.
This cost disparity is mitigated by the technology's potential for cost-effective solutions and predicted cost decrease with increased demand, making NB-IoT an appealing prospect for IoT deployments, particularly those in UAV-assisted systems. Furthermore, its narrow bandwidth approach improves spectrum utilization, making it especially ideal for applications that require enormous device connectivity, such as the growing smart city concept. Finally, the choice between NB-IoT and other IoT protocols is determined by a careful examination of specific IoT application requirements, which includes balancing factors such as power efficiency, cost-effectiveness, latency, coverage, and other critical parameters to align with the unique demands of the intended IoT deployment. In Fig.~\ref{fig:Comm}, we show the applicability of the different communication technologies in different networks.

% \subsubsection{Other Communication Technologies}
% In the realm of Unmanned Aerial Vehicles, specialized low-power wide-area network (LPWAN) technologies like ZigBee and SigFox offer distinct advantages and find applications in diverse scenarios. ZigBee operates as a high-level communication protocol, employing low-power digital radios for personal area networks. It strikes a balance between low data rates, extended battery life, and secure networking. For UAVs, ZigBee showcases its capabilities in various applications, notably demonstrated in a communication protocol where it facilitated image transmission between a UAV and a ground station using a Raspberry Pi 3 Model B board and the XBEE PRO S3B 915 MHz module embedded in a DJI Phantom 3 Standard UAV.

% On the other hand, SigFox serves as a global communication solution dedicated to the Internet of Things (IoT), utilizing unique technology to power its network. It excels in handling low data-rate messages efficiently, following a ``send and forget'' approach. This makes SigFox ideal for applications where minimal data usage and power efficiency are crucial. While specific UAV-centric applications or experiments leveraging SigFox are not widely documented, its potential in the UAV domain remains promising, emphasizing low power consumption and extended communication ranges. In conclusion, selecting between ZigBee and SigFox depends on aligning their strengths with the specific requirements of the intended IoT application, making informed choices for efficient and effective UAV deployments.

\begin{figure}[!t]
    \centering 
    \includegraphics[width=\columnwidth]{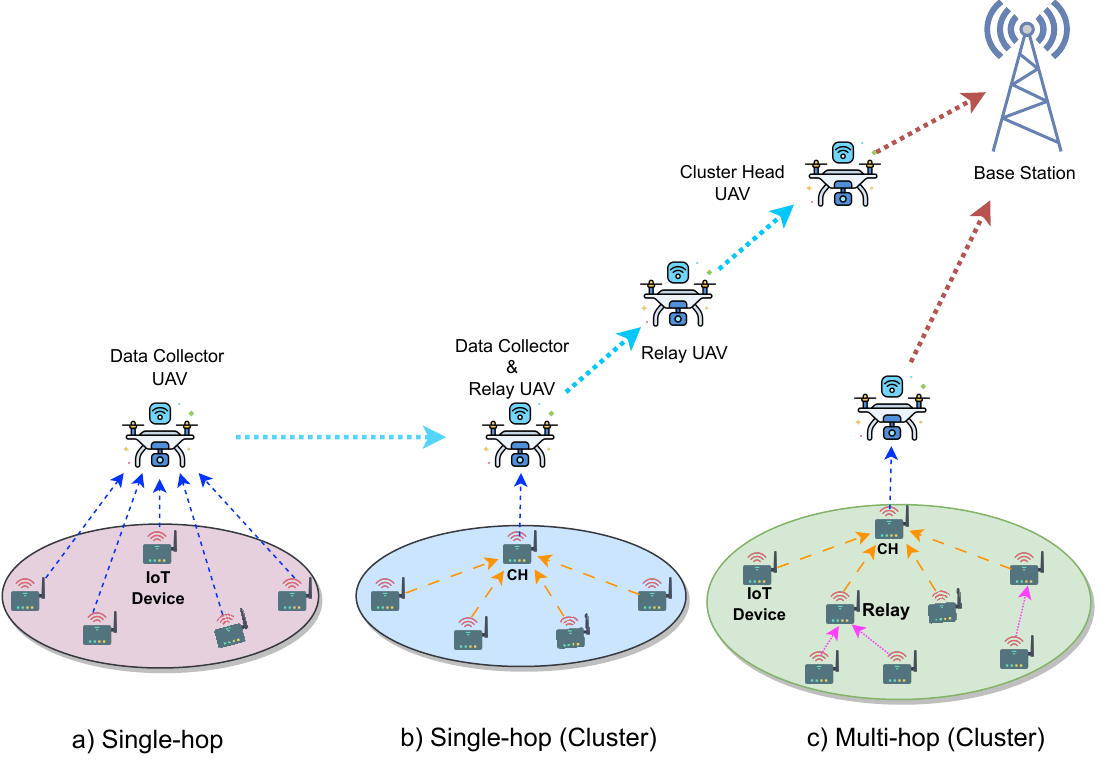}
    \caption{Different communication scenarios in UAV-based IoT Networks}    
    \label{fig_2:Network}
\end{figure}
\subsection{Categorization of UAV-IoT Network Topologies}

In Fig~\ref{fig_2:Network}, we show different communication network scenarios in UAV-assisted IoT Networks. Each of these networks is discussed as follows: 

\begin{itemize}
    \item Single-hop IoT network: In Fig.~\ref{fig_2:Network}(a), we show that all the IoT nodes individually transmit their information to the UAV that is in their communication range.
    \item Cluster head (CH) based IoT network: As shown in Fig.~\ref{fig_2:Network}(b), IoT nodes choose a CH and transmit their information to the chosen CH, who is responsible for further transmitting the collected data to its connected UAV. In this kind of network, IoT nodes are in the communication range of each other.
    \item Mulit-hop CH-based IoT network: It is similar to the CH-based IoT networks, but IoT nodes are not in the communication range of each other. They form multi-hop networks among themselves.
    \item Single-hop UAV network: In this network, the data collector UAV is in the communication range of the Base station so that the UAV can transmit to the Base station without any intermediate relay UAV.
    \item Multi-hop UAV network: Here, the data collector UAV is not in the communication range of the base station, so it should take help from the Relay UAV to transmit data to the base station. In this type of network scenario, some of the UAVs work just as Relay UAVs. They do not collect data from the IoT nodes. The UAV nearest to the base station can be considered as CH of the UAVs.   

\end{itemize}

\section{Overview of Data Collection Methods}
\label{sec:fundamentals2}
In this section, we discuss all the typical UAV deployment forms for data collection, as well as performance metrics, one by one as follows. %The node location distribution models for both deployment forms differ from those for traditional communication contexts (e.g., cellular). 

\subsection{Hovering-based UAV Data Collection}
In Hovering (aka static) UAV mode, the UAV remains above the selected data-collecting spot for a period of time to collect data. In scenarios where the terrestrial communication infrastructure is unavailable, damaged, or inadequate due to high user volume, strategic placement of UAVs becomes crucial. When focusing on a single ground receiver, the UAV should ideally be positioned in close proximity to optimize transmission if it's acting as a transmitter. Conversely, when functioning as a relay, determining the optimal position requires careful consideration of both the transmitter and receiver locations \cite{chen2017optimum}. 

In cases where multiple ground users need to cover, a critical design objective is required to ensure adequate coverage while minimizing costs, which could be related to power consumption or the number of UAVs needed. Achieving such objectives involves optimizing the UAV's three-dimensional (3D) position \cite{alzenad20173}. Additionally, in situations with multiple UAV transmitters, achieving comprehensive ground coverage can be framed as a ``disk covering problem''. The UAV's coverage area forms a circular disk on the ground, and the goal is to cover the target area with the fewest disks, thereby minimizing costs. This challenge has been extensively studied in the literature \cite{lyu2016placement}. In \cite{bushnaq2019aeronautical}, the entire area is divided into sub-regions, and the UAV, which is stationed in a fixed position over these sub-regions to collect data. In brief, hovering mode provides reliable data transfer at a greater energy cost.

\subsection{Flying UAV Data Collection}
In Flying (aka mobile) UAV data collection mode, the UAV efficiently acquires sensor data while decelerating as it approaches the data collecting spot, stopping data collection only when it leaves the communication range of the sensors. The ability to capture data while moving ensures that the data collection operation is completed quickly. However, if the volume of data is large, the UAV in this mode may struggle to meet the data-collecting timeliness requirement. The major goal of mobile UAV data collecting is to minimize flying duration frequently. According to research in UAV-assisted data collection, the best UAV speed is directly proportional to sensor energy and density but inversely proportional to data volume for upload. As a result, studies optimize parameters such as the horizontal distance across which sensors can transmit data to the UAV, UAV speed, and sensor transmit power to shorten UAV flying time during data collecting. This optimization strategy is obvious in works such as \cite{gong2018aviation}. \cite{zhan2019completion} focus on lowering the maximum data collection completion time across all UAVs in the domain of multi-UAV-enabled IoT. Their approach involves the joint optimization of UAV trajectories, wake-up scheduling, and sensor association, ensuring successful data upload by sensors within specified energy constraints and shedding light on this intricate trade-off relation while considering the constraint of limited energy for both UAVs and sensors \cite{zhan2019energy}.

\subsection{Hybrid UAV Data Collection}
The Hybrid (aka Collaborative) mode of data collection in UAV systems combines the characteristics of both static and mobile modes. In this mode, the UAV stays above a data collection point to gather data and continues to collect data as it flies to the next point until it is out of the communication range of the sensor \cite{poudel2021hybrid}. This mode is more energy-efficient than the static mode but less so than the mobile mode. The choice of data collection mode can be dynamically adjusted according to the energy consumption of the UAV. For instance, \cite{cui2022uav} proposed a UAV path planning method for data collection of fixed-point equipment in complex forest environments. The work proposed two-point path planning and multi-point path planning methods to maximize the amount of fresh information collected from ground-fixed devices in a complicated forest environment. The authors adopted a chaotic initialization and co-evolutionary algorithm to solve the two-point path planning issue considering the UAV's performance and environmental factors. Then, a UAV path planning method was designed based on simulated annealing for the multi-point path planning issue. A hybrid hovering points selection algorithm considering the limited battery of machine-type communication devices (MTCDs) and UAVs proposed in \cite{zhu2018energy}. The algorithm selects the hovering points of UAV with minimum power consumption of MTCDs and allows for dynamic adjustment of the data collection mode based on application scenarios, performance requirements, and energy constraints.

In both methods, the UAV can switch between static and mobile data collection modes based on the requirements of the data collection task. For example, it can start with mobile data collection to cover a large area quickly and then switch to static data collection for areas that require more detailed data gathering. %This approach combines the advantages of both modes, providing flexibility and efficiency in data collection.

% \begin{figure*}[!t]
%     \centering 
%    \includegraphics[width=.9\linewidth]{Image/Static_and_mobile.pdf}
%     \caption{Static and Mobile Data Collection}    
%     \label{fig_5:SnM}
% \end{figure*}

\subsection{Advantages and Disadvantages of Each Method}

\begin{table*}[t]
\caption{Comparison of different data collection schemes}
\label{DCM}
\scriptsize
\tiny
\setlength\extrarowheight{3pt}
\resizebox{\textwidth}{!}{%
\begin{tabular}{|c|p{2.5cm}|p{2cm}|p{2.5cm}|p{2cm}|p{2.5cm}|}
\hline \hline
\rowcolor[HTML]{8FEDDD} 
\textbf{Data Collection Mode} & \textbf{Scheme} & \textbf{AoI} & \textbf{Energy Efficiency} & \textbf{Flight Time} & \textbf{Throughput} \\ \hline \hline

\rowcolor[HTML]{EFEFEF}
 &
 
  UAV Trajectory Planning for AoI - Minimal Data Collection \cite{9892691} & Minimal & High & Medium & Depends on data size \\ 
\cline{2-6}
\rowcolor[HTML]{EFEFEF}
Static/Hovering &
  Energy-Constrained UAV Data Collection Systems: NOMA and OMA \cite{9447204} & Tailored to meet specific energy budgets at both the UAV and GNs & Tailored to meet specific energy budgets at both the UAV and GNs & Tailored to meet specific mission requirements & Maximized minimum UAV data collection throughput \\ 

\cline{2-6}
\rowcolor[HTML]{EFEFEF}
 &
  UAV-Aided Cooperative Data Collection Scheme for Ocean Monitoring \cite{9377450}& Tailored to meet specific mission requirements & Tailored to meet specific mission requirements & Tailored to meet specific mission requirements & Maximized network lifetime \\ 

\hline

 &
 AoI-Energy-Aware UAV-Assisted Data Collection \cite{9426899} & Minimized & Medium & Low & Depends on data size and the application \\ 

\cline{2-6}
Mobile/Flying &
  Joint Mobile Vehicle–UAV Scheme for Secure Data Collection in a Smart City \cite{huang2021joint} & Customized for security and efficiency & Customized for security and efficiency & Customized for security and efficiency & Designed for efficient, cost-effective, and secure data collection \\ 

% \cline{2-6}

%  &
%   UAV-Aided Cooperative Data Collection Scheme for Ocean Monitoring &
%   Depends on scheme3 &
%   Depends on scheme3 &
%   Depends on scheme3 &
%   Maximized network lifetime3 \\
\hline
% \rowcolor[HTML]{EFEFEF}
%  &
%   UAV Trajectory Planning for AoI-Minimal Data Collection &
%   Minimized1 &
%   High1 &
%   Medium1 &
%   Depends on data size1 \\
% \cline{2-6}
% \rowcolor[HTML]{EFEFEF}
%  &
%   AoI-Energy-Aware UAV-Assisted Data Collection &
%   Minimized4 &
%   High4 &
%   Medium4 &
%   Depends on data size4 \\
% \cline{2-6}
\rowcolor[HTML]{EFEFEF}
 &
  Hybrid Path Planning for Efficient Data Collection in UAV-Aided WSNs \cite{poudel2021hybrid}& Optimized for collision-free paths in emergency scenarios & Significantly improved compared to other schemes based on specific requirements & Significantly improved compared to other schemes based on specific requirements & Significantly improved compared to other schemes based on specific requirements \\ 

\cline{2-6}
\rowcolor[HTML]{EFEFEF}
 Hybrid/Collaborative &
 AUV-Assisted Data Gathering Scheme Based on Clustering and Matrix Completion\cite{9068243} & Improved data gathering efficiency in underwater wireless sensor networks (UWSNs) & Enhanced efficiency in UWSNs & Enhanced efficiency in UWSNs & Improved data gathering efficiency in underwater wireless sensor networks (UWSNs) \\ 

\hline \hline
\end{tabular}%
}
\end{table*}

In the realm of UAV systems, data collection is a critical aspect that can be approached in various ways, each with its own advantages and potential drawbacks. 
% \textcolor{red}{Do not hard code like this. Also table title is missing. \textbf{Table II}} 
Table ~\ref{DCM} compares various approaches under different data collecting modes based on different performance metrics. Static data collection schemes, such as  \cite{9892691} and  \cite{9447204}, are ideal for scenarios where minimizing the AoI is crucial or where maximizing the minimum UAV data collection throughput from Ground Nodes (GNs) is important, subject to the energy budgets at both the UAV and GNs. On the other hand, mobile data collection schemes like  \cite{9426899} and  \cite{huang2021joint} are beneficial in scenarios where the IoT devices are widely dispersed or when the area to be covered is large. These schemes are particularly effective in smart city environments where secure data collection is a priority. Hybrid or collaborative data collection schemes combine the characteristics of both static and mobile modes. Schemes such as  \cite{poudel2021hybrid} and \cite{huang2020auv} are more energy-efficient than the static mode but less so than the mobile mode. These schemes are particularly effective in emergency or underwater environments where efficient data gathering is crucial. Therefore, choosing an appropriate scheme requires careful consideration of various factors such as AoI, energy efficiency, flight time, throughput, and specific application scenarios. 

\subsection{Performance Metrics}
Understanding and evaluating the performance through metrics is paramount for reliable and efficient data collection. Specifically, within the context of UAV-based DC, numerous key metrics are employed to comprehensively gauge efficiency levels. These metrics are pivotal in enabling researchers and contributors to optimize DC performance and showcase the enhancements brought about by novel contributions in this domain. In this section, we delve into these fundamental metrics, shedding light on their significance and applications in assessing and enhancing UAV-assisted data collection.

\begin{enumerate}
    \item \textbf{Data Collection Completion Time (\(T\))}:
    This metric is represented as \(T = t_{\text{end}} - t_{\text{start}}\) and provides valuable insights into the overall duration taken for the UAV to complete the data collection process. A shorter completion time indicates higher efficiency and swift data gathering from sensors or designated areas \cite{nguyen20223d}.

    \item \textbf{Throughput (\(T_H\))}:
    Throughput is denoted as \(T_H = \frac{D}{T}\) and quantifies the rate of successful data transmission over a specified time period. It reveals the system's ability to handle data transfer efficiently, considering both the amount of data transmitted (\(D\)) and the time taken (\(T\)). Higher throughput signifies a more efficient data collection process  \cite{nguyen20223d}.

    \item \textbf{Energy Efficiency (\(E\))}:
    Energy efficiency is expressed as \(E = \frac{D}{E_c}\), which measures the amount of data successfully transmitted per unit of energy consumed (\(E_c\)). Optimizing energy efficiency is critical to prolonging UAV mission durations and reducing the system's overall energy consumption  \cite{nguyen20223d}.

    \item \textbf{Age of Information}:
    AoI is calculated as \(AoI = t_{\text{new}} - t_{\text{old}}\), represents the `\textit{freshness}' of the collected information. It helps in understanding how current the data is, providing crucial insights for time-sensitive applications. A lower AoI indicates more up-to-date information \cite{sarkar2022framework}.

    \item \textbf{Coverage (\(C\))}:
    Coverage, assessed as \(C = \left(\frac{a}{A}\right) \times 100\%\), gauges the geographical area covered by the UAV during its flight. It provides an understanding of the spatial efficiency of the data collection process, aiding in optimizing flight trajectories and data collection strategies \cite{9779853}.

    \item \textbf{Reliability (\(R\))}:
    Reliability measures the probability that the UAV system will perform its required functions under specified conditions for a defined period. It's crucial for ensuring consistent and accurate data collection, especially in critical applications \cite{9817083}.
    \begin{equation}
        R = \frac{\textit{Number of Successful Data Transmissions}}{\textit{Total Number of Data Transmission Attempts}} \nonumber
    \end{equation}
    %\(R = \frac{\text{Number of Successful Data Transmissions}}{\text{Total Number of Data Transmission Attempts}}\)

    \item \textbf{Latency (\(L\))}:
    Latency is represented by \(L = t_{\text{transmit}} - t_{\text{send}}\), which indicates the delay between issuing the instruction to transmit data and the actual transmission. Lower latency is essential for real-time applications, ensuring timely data delivery and decision-making \cite{9817083}.
\end{enumerate}

%\subsection{Challenges in Data Collection}

% \section{Data Collection Techniques in UAV-based IoT Networks}
% \label{sec:data-collection}

% \subsection{Case Studies and Examples}

\section{Reliable data collection techniques in UAV-assisted IoT}
\label{sec:reliable}
This section focuses on three key concerns: ensuring accurate data, enhancing network connectivity, and securing data. We begin with error correction coding and re-transmission methods to improve data accuracy. Next, we explore specialized routing protocols to enhance network connectivity and wireless channel access mechanisms for UAV-IoT networks.  We also briefly discuss data security and privacy measures in such UAV-IoT interconnected environment.
\subsection{Data Accuracy and Consistency}
Data accuracy and consistency are of paramount importance in UAV-assisted IoT networks. Accurate data is crucial for making reliable decisions, such as in intelligent transportation systems where UAVs gather data from a network of sensing devices. Consistent data allows for efficient resource allocation, ensuring optimal performance of the network. Furthermore, accurate and consistent data is essential for maintaining the security of the network and mitigating risks associated with malicious data collectors. In emergency situations, the accuracy and consistency of data collected by UAVs can significantly impact the effectiveness of emergency communications. Here are some key aspects to consider:

\textbf{Bit Error}: Bit errors during data transmission in UAV-assisted IoT networks can detrimentally impact data accuracy, often stemming from interference, signal fading, and noise. To mitigate these issues, error detection and correction methods are crucial. Forward Error Correction (FEC) codes stand out as a technique capable of detecting and rectifying bit errors without necessitating data re-transmission. 

Bit error tolerance in UAV networks is essential for ensuring dependable and effective communication.  Advanced Tail-Biting Turbo-Hadamard Codes are a proposed solution for ensuring reliable communication in UAV networks, particularly considering short data packets in UAVs \cite{dong2022efficient}. These are constructed by concatenating multiple tail-biting convolutional-Hadamard codes in parallel are used in the scheme. This strategy is intended to reduce communication latency. These codes are incorporated into an anti-jamming frequency hopping (FH) communications system to ensure the reliability of UAV communications links in a hostile jamming environment. The component convolutional-Hadamard codes are designed using tail-biting, and a corresponding decode algorithm is implemented in the UAV hardware platform. The anti-jamming efficacy of this method is superior to that of conventional concatenated codes \cite{dong2022efficient}. Balanced performance strategy seeks to simultaneously decrease UAV network energy consumption, transmission delays, and bit error rate \cite{khalil2022balanced}. A UAV is equipped with a wireless bidirectional relay to increase transmission throughput and extend network coverage. The disadvantage of having a relay is the data transmission delay induced by the UAV's movement. This delay results in an increase in total energy consumption, which further degrades bit error rate performance, particularly at high SNR levels \cite{khalil2022balanced}. Utilising a multi-objective algorithm, an adaptive optimal energy allocation strategy is generated based on the equilibrium between energy consumption and transmission delays.

In decentralised UAV networks, the Raptor Coded Random Access (RCRA) protocol is essential. The RCRA programme implements a three-tiered strategy to improve the overall performance and efficacy of these complex systems \cite{9409942}. At first, it optimises the number of slots, and the second aspect of the RCRA scheme is the use of error-correcting codes as a pre-coding measure before random access. This step involves fine-tuning the access probability to achieve the lowest Bit-Error Ratio (BER). The third component of the RCRA scheme is an innovative approach to replacing empty slots. This method significantly improves the overall performance of random access systems by establishing correlations between two consecutive positions. This scheme significantly reduces the Block-Error-Ratio (BLER) and the BER, enhancing the robustness and dependability of decentralised UAV networks \cite{9409942}.

    % Addressing these bit error concerns, a joint optimization framework known as the Joint Optimization of Deployment and Trajectory (JOLT) \cite{9802633} has been proposed. This approach employs the Adaptive Whale Optimization Algorithm (AWOA) for UAV deployment optimization, overcoming limitations observed in the whale optimization algorithm, such as slow convergence and susceptibility to local optima. The AWOA achieves this through adaptive adjustment of weighting functions and differential variation disturbances, enhancing searching capabilities in later stages and mitigating the risk of falling into local optima. Furthermore, to optimize UAV trajectory and minimize bit errors, an Elastic Ring Self-Organizing Map (ERSOM), a variant of the Self-Organizing Map (SOM), has been introduced. ERSOMs build on the principles of SOMs, employing neural networks to reduce the dimensionality of high-dimensional data. They achieve this by flexibly moving the reference vector to reconstruct the distribution of high-dimensional data, providing more precise visualization. This flexibility allows for a nuanced representation of the data, ultimately improving the accuracy of the collected information.
    
 \textbf{Re-Transmission:} In certain scenarios, data packets might require re-transmission due to errors or losses during transmission, impacting data timeliness and freshness. When bit errors are detected in received packets, a re-transmission process is initiated to guarantee reliable data transmission in UAV networks. The Automatic Repeat Request (ARQ) protocol, which includes acknowledgement (ACK) and timing strategies, is frequently used in these re-transmission procedures. The receiver transmits an ACK frame to indicate the effective reception of the sender's packet. If a received packet contains bit errors, the receiver sends a Negative-Acknowledgment (NACK) frame to inform the sender of the unsuccessful reception. ARQ in UAV consists of Stop-and-Wait (SW), Go-back-N (GBN), and Selective Repeat (SR). In SW protocol, the originator cannot transmit additional packets without an ACK \cite{aljehani2020uav, 9989385}. In contrast, GBN allows the emitter to continue transmitting packets without an individual packet ACK up to a window size limit.

To address this challenge, a layered UAV swarm network architecture has been proposed \cite{8599015}. This architecture integrates a low latency routing algorithm (LLRA), designed based on partial location information and network connectivity. By this approach, the packet delivery ratio is enhanced, and the link average delay is reduced, ensuring efficient re-transmissions and maintaining data consistency. LLRAs are specifically designed to respond with minimal delay swiftly. They continuously consume market data from target exchanges, enabling rapid decision-making in microseconds and timely dispatch of order messages to multiple exchanges. Using this approach, data packets are forwarded while circumventing high-risk areas, minimizing the risk of potential attacks on user data.

Some cutting-edge research in the field of UAV network re-transmission includes the following. \cite{aljehani2020uav} presents an innovative strategy for establishing a secure heterogeneous network environment with the assistance of UAVs using an IP prototype that is seamless. It emphasises the significance of re-establishing communication tunnels between UAVs and end devices (EDs) during their traversal of heterogeneous networks to maintain uninterrupted communication. This method guarantees the uninterrupted and secure operation of UAV communication and control systems, especially when network conditions change. Although this approach is not explicitly focused on re-transmission, it indirectly contributes to data reliability by enabling seamless connectivity even when network conditions change. \cite{sayeed2020efficient} offers a comprehensive strategy for maximising throughput in terrestrial networks augmented by UAVs. The primary objective is to minimise delays and packet losses by optimising UAV trajectories and reinforcing congested nodes and transmission channels. While this paper does not explicitly discuss re-transmission techniques, its primary goal of minimising packet loss contributes inherently to reducing the need for re-transmissions. In turn, this improves the dependability of data transmission in UAV networks. In the domain of re-transmission techniques within UAV networks, two research papers contribute significantly to data transfer reliability and security. \cite{9613552} underscores the importance of secure downlink data transmission through NOMA and optimising UAV trajectories and resources. While not explicitly addressing re-transmission, the paper indirectly enhances data transfer reliability by minimizing errors. Conversely, \cite{9143289} focuses on security measures to mitigate eavesdropping and enhance privacy protection in aerial-ground transmission. Although not directly addressing re-transmission, it indirectly contributes to error reduction, reducing the reliance on re-transmission for maintaining data transfer reliability. %The key difference lies in their primary emphasis, with the former paper highlighting optimization techniques and the latter paper emphasizing security measures within UAV networks \cite{9613552, 9143289}.

% \textcolor{red}{Need to add more papers on Bit-error and Re-transmission. Check \cite{9647007}}
\color{black}
\subsection{Network Connectivity}

In terms of network connectivity, UAV networks exhibit a wide range of dynamics, ranging from slow-moving to extremely fast operations. The nodes are subject to intermittent failures or power limits, necessitating rapid replacements to provide an uninterrupted and dependable connection. However, maintaining reliable connectivity in UAV networks is frequently hampered by disturbances in links caused by the location of UAVs and base stations. These lines must also struggle with high bit error rates due to interference or natural obstructions. Striking a delicate balance between reliable connectivity and power efficiency is critical, especially considering the different dependability requirements inherent in UAV networks. For example, delivering essential seismic data necessitates a perfect, dependable transport mechanism to ensure vital information reaches its intended destinations precisely and on time. When sending photographs and videos of the earthquake, a somewhat lower level of reliability may be tolerated, but strict latency and jitter standards must be met. In such a complicated environment, bandwidth requirements for different data types, such as voice, data, or video, are distinct, reflecting the multiple needs of UAV networks. These networks incorporate the core elements of typical mobile wireless networks while adding complexity. Navigating this complexity is a considerable issue, as nodes' mobility, network partitioning, intermittent links, restricted resources, and changing QoS requirements are critical in providing dependable and efficient network connectivity in UAV networks.

%\textcolor{blue}{\cite{7317490} -- contains everything about Routing and MAC} \textcolor{red}{ Follow this paper and add ref related to UAV}
    \subsubsection{\textbf{Routing} } In the pursuit of enhanced communication protocols for aerial networks, researchers have extensively studied the protocols used in Mobile Ad-Hoc Networks (MANETs) and Vehicular Ad-Hoc Networks (VANETs) due to the apparent similarities with UAV networks. Despite the search for improved protocols continuing in these environments, multi-UAV networks present a unique set of requirements to consider, including managing mobility patterns, nodes' localization,  link control, nodes' additions and removals, diverse application areas, and specific QoS needs due to power constraints.

Addressing the distinct challenges of UAV networks, modifications have been proposed for MANET protocols. However, the peculiar issues faced by UAV networks necessitate the development of new routing algorithms to ensure reliable communication among UAVs \cite{mozaffari2019tutorial} and between UAVs and the control centre(s). To assess their applicability in UAV Networks and their potential usefulness, we will explore several networking protocols based on the widely recognized classification as follows:

%Routing Protocols for UAVs:
%Routing protocols for UAVs are designed to enable efficient and reliable communication and data transfer among UAVs and between UAVs and ground control stations. These protocols need to account for the dynamic and often ad-hoc nature of UAV networks. Several routing protocols have been proposed and adapted for UAV networks, and the choice of protocol may depend on the specific application, network size, mobility patterns, and other factors. Here are some routing protocols commonly used or proposed for UAVs:

\textbf{Proactive Routing Protocols : }
Proactive routing protocols maintain up-to-date routing information at all times, regardless of whether it's needed or not. This employs tables in their nodes to store all of the routing information of other nodes or nodes in a certain network region. When the topology changes, the tables must be updated. The key advantage of proactive routing is that it contains the most up-to-date route information. These are suitable for scenarios where network topology changes are relatively slow and predictable.

\textbf{Optimized Link State Routing} (OLSR) falls under this category and maintains a proactive routing table. Initially, routes are determined and continually updated through topology information exchange among nodes, using a flooding strategy to broadcast link-state costs of neighbouring nodes. This facilitates efficient routing by applying a shortest path algorithm to select the optimal next hop towards various destinations, ensuring effective communication within the network \cite{7118984}. In the dynamic context of UAV networks, rapid changes in node locations and links pose challenges for the traditional OLSR approach, resulting in an increased number of control messages and elevated overhead. To optimize routing in this context, OLSR implements Multipoint Relays (MPRs), designating specific nodes to forward control traffic. This strategic selection significantly reduces the transmission load and minimizes control message overhead, enhancing the efficiency and streamlining the routing process in UAV networks, which are often resource-constrained and rapidly changing \cite{7118984}. Variants of OLSR further tailor the algorithm to address specific challenges and network dynamics. Global State Routing (GSR) optimizes by restricting update messages to intermediate nodes, aiming for improved efficiency. Fisheye State Routing (FSR) reduces overhead by providing more frequent, smaller updates to nearby nodes within a designated scope. However, this may introduce inaccuracies, particularly in mobile node networks. Source-Tree Adaptive Routing (STAR) reduces overhead by conditioning update dissemination, albeit with increased memory and processing requirements in large and mobile networks \cite{shenbagapriya2014survey}. Another variant, the Distance Routing Effect Algorithm for Mobility (DREAM), leverages GPS-derived geographical coordinates to optimize routing efficiency, allowing for proportional updates based on node velocity and reducing overall overhead. These variants demonstrate the flexibility and adaptability of OLSR, catering to diverse network scenarios and ensuring efficient communication within ad-hoc networks \cite{peters2011geographical}.

\textbf{Reactive (On-Demand) Routing Protocols: }
It establishes routes on-demand when a data packet needs to be sent. They are suitable for scenarios with dynamic and unpredictable network topology changes. Reactive protocols, due to their on-demand nature, are bandwidth-efficient as they avoid periodic messaging. However, the route discovery process might introduce high latency since finding routes can be time-consuming. Reactive protocols are categorized into source routing and hop-by-hop routing. In source routing, data packets contain the complete source-to-destination address, allowing intermediate nodes to forward packets based on this information. While this approach eliminates the need for periodic beaconing and ensures connectivity, it does not scale well as route failure probability and header size increase with network size, resulting in higher overhead. On the other hand, hop-by-hop routing involves each data packet carrying only the destination and next hop addresses. Intermediate nodes maintain routing tables to forward data, allowing adaptability to the changing environment. However, this strategy requires each intermediate node to store and maintain routing information for active routes, and nodes need to be aware of neighbouring nodes through beaconing messages, presenting a potential overhead in the UAV network \cite{6477825}.

\textbf{Ad-Hoc On-Demand Distance Vector}  (AODV)  is a reactive routing protocol designed for mobile ad-hoc networks, utilizing a hop-by-hop approach. It dynamically adjusts to changing link conditions, providing advantages such as low processing and memory overhead, as well as low network utilization. AODV optimizes routing traffic by building routes on demand. However, it does face delays during route construction and potential bandwidth consumption issues during route discovery, especially with growing network size and in the presence of link or node failures in UAV networks \cite{fu2007mesh}. For improved routing in aerial applications, a combined protocol known as Reactive-Greedy-Reactive (RGR) has been proposed, integrating Greedy Geographic Forwarding (GGF) and reactive routing mechanisms. RGR leverages UAV location data and reactive end-to-end paths, showcasing enhanced efficiency, particularly in search missions, with superior performance in delay and packet delivery ratio compared to existing protocols like AODV \cite{shirani2012combined}.

\textbf{Dynamic Source Routing} (DSR) is primarily intended for mobile node multi-hop wireless mesh ad-hoc networks. It enables networks to self-organize and self-configure without relying on current infrastructure. DSR operates on-demand and immediately responds to changes in existing routes. Its ``route discovery'' and ``route maintenance'' technologies allow nodes to identify and maintain routes to numerous destinations, allowing for multiple route selection and applications such as load balancing. It also ensures loop-free routing \cite{johnson2007dynamic}. However, building new routes in UAV networks using DSR can be time-consuming because each packet must transmit the addresses of all nodes from source to destination, making it less suitable for large or dynamically changing networks \cite{brown2004ad}.

\textbf{Geographic Routing Protocols} (GRP) plays a vital role, leveraging the geographic locations of nodes. A common method employed in geographic routing is greedy forwarding. Greedy forwarding involves each node directing the message to the nearest neighbour to the destination based on their geographic proximity, utilizing only local information. However, challenges may emerge if the message gets trapped in a local minimum during this process, causing a halt in progress. To tackle this, recovery mechanisms like face routing come into play, finding an alternative path to resume packet forwarding and ensure successful delivery in UAV networks \cite{bose1999routing}.

\textbf{Greedy Perimeter Stateless Routing} (GPSR) is a position-based routing mechanism frequently employed in UAV Ad-hoc networks. GPSR's data forwarding process is bifurcated into two modes: greedy forwarding and perimeter forwarding \cite{9463341}. In the greedy forwarding mode, packets are forwarded to the neighbouring node that is geographically closest to the destination. This process continues until the packet reaches its destination or encounters a node without any neighbours closer to the destination \cite{samundiswary2010secured}. When greedy forwarding fails, GPSR switches to perimeter forwarding. However, the traditional approach of using the right-hand rule to identify the next hop node in perimeter forwarding can lead to detours. This can increase overhead in UAV networks with rapidly changing topologies \cite{9463341}. To mitigate this issue, an optimization algorithm based on a game model has been proposed. This model considers the forwarding angle and distance to design a scoring function and revenue model for bidding nodes. The optimal next-hop node is selected by comparing the quality attributes and bidding prices of multiple bidding nodes \cite{9463341}.

\textbf{Hybrid Routing Protocols} (HRP) combines features of both proactive and reactive protocols to adapt to changing network conditions. They use proactive techniques for some parts of the network and reactive techniques for other parts. The large latency of the initial route discovery stage in reactive routing protocols can be decreased by employing HRP, as can the overhead of control messages in proactive routing systems. It is especially well-suited for large networks, and the network is divided into zones, with Interzone routing performed in advance and Interzone routing handled reactively. Based on network characteristics, hybrid routing alters its strategy.

\textbf{Zone Routing Protocol} (ZRP) exemplifies this hybrid approach by segmenting the network into zones and applying different routing strategies within and between these zones \cite{6386782}. Within each zone, ZRP employs proactive routing, maintaining up-to-date routing information for the zone. This proactive strategy helps to minimize the initial route discovery delay typically associated with reactive routing. For communication between zones, ZRP switches to a reactive routing strategy. This approach helps to decrease control traffic, a common issue with proactive routing. When a source node needs to send data to a destination outside its local zone, it initiates a route discovery process. The combination of proactive and reactive strategies allows ZRP to adapt to changing network conditions. This makes it particularly suitable for networks with dynamic topologies, such as UAV networks \cite{6386782}.

\textbf{Opportunistic Routing Protocols} (ORP) take advantage of multiple forwarding opportunities to increase the chances of data delivery, which can be useful in challenging network conditions.  From the standpoint of data communication, this protocol constructs a data forwarding priority model and sets data transmission rules in conjunction with a node activity range division method \cite{8966343}. These protocols are very relevant in UAV-based IoT systems because network topology might change quickly due to UAV mobility. Opportunistic routing techniques can assure reliable and efficient communication even in highly dynamic networks by taking advantage of every feasible opportunity for data forwarding.

\textbf{Spray and Wait} is a prominent opportunistic routing protocol that operates in two distinct stages. The spray stage and the wait stage. During the spray stage, a predetermined number of copies of the message are disseminated throughout the network. The quantity of copies is influenced by various factors, including the velocity and remaining energy of nodes. Following the dissemination, the protocol transitions to the wait stage. In this stage, each node holding a copy of the message waits until it encounters the destination node directly. Spray and Wait has been modified and optimized in numerous ways to enhance its performance. For example, the Sharing Spray and Wait routing algorithm determines the next node and the number of copies a node can deliver based on the duration of message carriage and the probability of message delivery \cite{5705251}. Another variant, Space-Aware Spray and Transfer Routing (SSTR) takes into account both the temporal and spatial information of mobile nodes. This helps nodes to select more competent nodes for further forwarding of packet copies \cite{10285692}. These modifications make Spray and Wait a versatile and resilient protocol for opportunistic routing in various network conditions.

\textbf{Delay-Tolerant Networking  Protocols} (DTN)-based routing protocols are designed for networks with intermittent connectivity. They store and forward data when network connectivity is sporadic. \textbf{Bundle Protocol} (BP) is a key component of the Delay-Tolerant Networking (DTN) architecture. It operates as an overlay on the top of transport protocols like TCP or UDP or lower-layer protocols like Bluetooth and Ethernet. The Bundle Protocol provides a unified interface for applications while allowing the underlying network to handle disruptions and long delays. This is particularly useful when network connectivity is sporadic or unreliable \cite{mcmahon2009delay}. In the context of UAV networks, BP can be particularly useful in disrupted environments where end-to-end paths between communication sources and destinations may not exist. This covers many applications, including transportation, traffic control, surveillance, border patrolling, search and rescue, disaster management, wireless network connectivity, smart agriculture and forestry, and remote immersion via mobile virtual reality \cite{al2021uav}. BP uses a ``store and forward'' mechanism, which means that data is stored at a node until it can be forwarded to the next node. This is particularly useful when network connectivity is sporadic or unreliable \cite{mcmahon2009delay}.

\textbf{Energy-Efficient Routing Protocols} (EERP)  are designed to minimize energy consumption during data transmission, which is crucial for UAVs with limited battery capacity. These protocols focus on optimizing the use of network resources, such as the energy of nodes and the bandwidth of communication links. For instance, an energy-efficient multilevel secure routing (EEMSR) protocol in IoT networks has been proposed \cite{9581300}. This protocol uses a cluster-based multi-hop routing approach to reduce the high communication overhead due to the scalability of IoT networks. In \cite{9411725}, an energy-efficient trajectory optimization scheme for UAV-assisted IoT networks has been proposed. In such networks, a single UAV is powered by both solar energy and charging stations (CSs), resulting in sustainable communication services while avoiding energy outages. Energy-Efficient Routing Protocols provide reliable communication in challenging network conditions while minimizing energy consumption. They are particularly useful in scenarios where the network nodes, such as UAVs, have limited battery capacity.

\textbf{Energy-Efficient AODV}  (EE-AODV) routing protocol is a variation of the AODV routing protocol that has been built primarily to improve energy efficiency in UAV networks \cite{tamizharasu2023intelligent}. The route discovery procedure in classic AODV can use much energy because it includes broadcasting route request packets across the network. EE-AODV includes energy awareness in the route discovery process to overcome this issue. The route request packets in EE-AODV include information about the nodes' energy condition. The nodes then use this information to make more energy-efficient routing decisions. For example, a node may choose to forward a packet through a route that is somewhat longer but contains nodes with greater energy levels. EE-AODV also provides algorithms for balancing energy usage among nodes and avoiding the overuse of certain nodes. This can help to extend the network's overall lifespan \cite{tamizharasu2023intelligent}.

\textbf{Hierarchical Routing Protocols} (HRP) divide the network into clusters or layers to manage the complexity and improve scalability.
HRP employs a layered approach to data routing within a network. These protocols segment the network into various hierarchical levels, each with its own distinct routing strategy. For instance, in a two-tier HRP, the lower tier might implement a proactive routing strategy while the upper tier employs a reactive routing strategy. This dual-strategy approach allows the protocol to balance the quick response time of proactive strategies with the low overhead of reactive strategies \cite{zhang2019iot}.
In the context of UAV networks, HRP can be particularly beneficial. A low latency routing algorithm (LLRA) is designed based on partial location information and the connectivity of the network architecture \cite{zhang2019iot}. They are especially useful in scenarios where the network topology is subject to rapid changes or is highly diverse.

\textbf{Hierarchical AODV} (HAODV)  is a variant of the AODV routing protocol. It’s designed to improve the scalability of Wireless Mesh Networks (WMNs). The Directional Hierarchical AODV (DH-AODV) routing protocol takes advantage of the existing fields of the control packets in the AODV to reduce the load on the network’s bandwidth and quickly detect route breakage. In \cite{7381367}, the DH-AODV protocol is designed, implemented, and evaluated using Qualnet Simulator. The simulation results indicate that this enhanced routing protocol can reduce the end-to-end delay of data packets by $47$ per cent, improve the throughput by $11.2$ per cent, and decrease the packet loss rate by $20$ per cent less than the standard AODV. This makes HAODV a more efficient and robust routing protocol for WMNs, which can be considered as a key technology for next-generation wireless networking. However, field trials and experiments prove that the performance of WMNs is still below expectations. Therefore, several challenging research issues need to be determined.

\textbf{Software-Defined Networking} (SDN) for UAVs applied to UAV networks to provide dynamic and centralized control over routing and network management. This allows for more flexibility and adaptability in the network, which is particularly beneficial in environments with high mobility and rapid changes in network topology, such as those involving UAVs \cite{8406959}. SDN separates the control and data planes, providing complete network programmability \cite{pasandideh2022topology}. This separation allows the SDN controller to make intelligent, optimal decisions based on the information it gathers from the UAV network itself. Moreover, an SDN framework for a UAV backbone network has been proposed, which includes a monitoring platform in the SDN controller to manage and analyze information from UAV networks effectively. Based on the analysis results, a load balancing algorithm is proposed to maintain a desirable network service \cite{8406959}. However, implementing SDN in UAV networks also presents several challenges, such as security, adaptability, interference management, interoperability, and lack of standardization. Despite these challenges, the combination of SDN and UAV-enabled networks is becoming essential in the 6G networking context and can offer more flexibility, scalability, reliability, and efficient connectivity than conventional approaches \cite{10274880}. Each category has specific advantages and is suitable for different use cases and network scenarios. The choice of routing protocol will depend on the unique requirements of the UAV network and the nature of the mission or application. We briefly summarised the above-mentioned routing schemes in Table~\ref{table_routing}.

\begin{table*}
\caption{Routing Protocols for UAVs: Advantages and Disadvantages} \label{table_routing}
\scriptsize
\centering
\setlength{\extrarowheight}{2pt} 
\begin{tabular}{|c|c|c|c|}
\hline  \hline
\rowcolor[HTML]{8FEDDD} 
\textbf{Category} & \textbf{Protocol} & \textbf{Advantages} & \textbf{Disadvantages} \\
\hline \hline
\rowcolor[HTML]{EFEFEF} 
\multirow{2}{*}{Proactive} & OLSR & 
\begin{minipage}{0.3\textwidth}
\begin{itemize}
  \item Up-to-date routes.
  \item Reduced route discovery delay.
  \item Low routing overhead.
\end{itemize}
\end{minipage} & 
\begin{minipage}{0.3\textwidth}
\begin{itemize}
  \item High control overhead.
  \item Not suitable for highly dynamic networks.
  \item Limited scalability.
\end{itemize}
\end{minipage} \\
\hline

\multirow{2}{*}{Reactive} & AODV & 
\begin{minipage}{0.3\textwidth}
\begin{itemize}
  \item Low control overhead.
  \item Efficient for smaller networks.
  \item Quick route discovery.
\end{itemize}
\end{minipage} & 
\begin{minipage}{0.3\textwidth}
\begin{itemize}
  \item Route discovery latency.
  \item Inefficient for large networks.
  \item May suffer from routing loops.
\end{itemize}
\end{minipage} \\
\cline{2-4}

& DSR & 
\begin{minipage}{0.3\textwidth}
\begin{itemize}
  \item Source routing flexibility.
  \item No need for route tables.
  \item Low routing overhead.
\end{itemize}
\end{minipage} & 
\begin{minipage}{0.3\textwidth}
\begin{itemize}
  \item High overhead in large networks.
  \item Increased packet header size.
  \item May lead to high latency.
\end{itemize}
\end{minipage} \\
\hline
\rowcolor[HTML]{EFEFEF} 
\multirow{2}{*}{Geographic} & GPSR & 
\begin{minipage}{0.3\textwidth}
\begin{itemize}
  \item Efficient for location-based routing.
  \item Reduced control overhead.
  \item Suitable for data dissemination.
\end{itemize}
\end{minipage} & 
\begin{minipage}{0.3\textwidth}
\begin{itemize}
  \item Dependency on accurate node location.
  \item Inefficient in non-geographic scenarios.
  \item May suffer from local minimums.
\end{itemize}
\end{minipage} \\
\hline

\multirow{2}{*}{Hybrid} & ZRP & 
\begin{minipage}{0.3\textwidth}
\begin{itemize}
  \item Combines proactive and reactive features.
  \item Efficient for varying network conditions.
  \item Low routing overhead.
\end{itemize}
\end{minipage} & 
\begin{minipage}{0.3\textwidth}
\begin{itemize}
  \item Increased complexity.
  \item May require more resources.
  \item Not well-suited for extremely dynamic networks.
\end{itemize}
\end{minipage} \\
\hline

\rowcolor[HTML]{EFEFEF} 
\multirow{2}{*}{Opportunistic} & Spray and Wait & 
\begin{minipage}{0.3\textwidth}
\begin{itemize}
  \item Increased delivery probability.
  \item Suitable for intermittent connectivity.
  \item Adaptive to changing network conditions.
\end{itemize}
\end{minipage} & 
\begin{minipage}{0.3\textwidth}
\begin{itemize}
  \item Higher overhead due to replication.
  \item Delay in delivery for some packets.
  \item Potential inefficient resource usage.
\end{itemize}
\end{minipage} \\
\hline

\multirow{2}{*}{DTN} & Bundle Protocol & 
\begin{minipage}{0.3\textwidth}
\begin{itemize}
  \item Handles disrupted environments.
  \item Effective in delay-tolerant scenarios.
  \item Improved data delivery predictability.
\end{itemize}
\end{minipage} & 
\begin{minipage}{0.3\textwidth}
\begin{itemize}
  \item Delayed data delivery.
  \item Complex routing decisions.
  \item May not be suitable for real-time applications.
\end{itemize}
\end{minipage} \\
\hline

\rowcolor[HTML]{EFEFEF} 
\multirow{2}{*}{Energy-Efficient} & EE-AODV & 
\begin{minipage}{0.3\textwidth}
\begin{itemize}
  \item Low energy consumption.
  \item Prolongs UAV battery life.
  \item Enhanced efficiency for energy-aware networks.
\end{itemize}
\end{minipage} & 
\begin{minipage}{0.3\textwidth}
\begin{itemize}
  \item Reduced network performance.
  \item Limited scalability in larger networks.
  \item May not adapt well to varying energy levels.
\end{itemize}
\end{minipage} \\
\hline

\multirow{2}{*}{Hierarchical} & HAODV & 
\begin{minipage}{0.3\textwidth}
\begin{itemize}
  \item Improved network scalability.
  \item Better resource management.
  \item Reduced routing overhead within clusters.
\end{itemize}
\end{minipage} & 
\begin{minipage}{0.3\textwidth}
\begin{itemize}
  \item Added complexity.
  \item Latency in inter-cluster routing.
  \item Cluster management overhead.
\end{itemize}
\end{minipage} \\
\hline

\rowcolor[HTML]{EFEFEF} 
\multirow{2}{*}{SDN for UAVs} & SDN-based & 
\begin{minipage}{0.3\textwidth}
\begin{itemize}
  \item Centralized control.
  \item Dynamic routing management.
  \item Improved adaptability.
\end{itemize}
\end{minipage} & 
\begin{minipage}{0.3\textwidth}
\begin{itemize}
  \item Requires compatible infrastructure.
  \item Potential single point of failure.
  \item Increased control plane traffic.
\end{itemize}
\end{minipage} \\
\hline \hline
\end{tabular}
\end{table*}

\color{black}

\subsubsection{\textbf{Medium Access Control} (MAC)}
 MAC protocols are crucial in UAV networks. They not only affect the system performance but also the energy efficiency in battery-powered sensor nodes \cite{8718663}. Numerous mission-critical applications, including surveillance, search and rescue, environmental monitoring, and package delivery, are frequently supported by these networks. There may be a requirement for many UAVs to connect with ground stations, other UAVs, or IoT gadgets. Given that the wireless channel is shared, a strong MAC protocol is necessary to prevent interference and collisions. Numerous MAC protocols with various goals have been developed for use in UAV-aided Wireless Sensor Networks (UWSNs).

\textbf{Random Access:} It is one of the MAC protocols by which UAVs can access the channel at any time when they are prepared to transmit their data. However, such random access might cause collisions. In order to mitigate this, UAVs monitor the channel and postpone transmission if they notice action already in progress. To increase channel efficiency, the Carrier Sense Multiple Access/Collision Avoidance (CSMA/CA) method is commonly used \cite{arafat2019mac}. In CSMA/CA, UAVs monitor the channel and defer transmission if they detect ongoing activity. UAVs employ \textit{back-off} mechanisms to reduce further the possibility of collisions \cite{klein2011bps}. When a UAV detects that the channel is occupied, it waits a random amount of time before attempting to transmit. After transmitting data, UAVs anticipate acknowledgement frames from the receiver, which indicate successful data reception. If the UAV does not receive an acknowledgement, it may re-transmit the data. In situations where multiple UAVs attempt transmission simultaneously (known as a collision), contention resolution mechanisms help manage conflicts and enable UAVs to reattempt transmission. Random access MAC enables UAVs to transmit data on demand, making it suitable for dynamic and unforeseeable mission scenarios. It can manage various UAVs in the network without requiring a fixed schedule, making it adaptable to changing mission specifications. UAVs can act autonomously without centralised control, enhancing their resilience and adaptability. Random access MAC, when combined with CSMA/CA and \textit{back-off} mechanisms, can optimise channel utilisation and reduce contention. Collisions can reduce network efficacy and increase latency, making effective contention resolution mechanisms crucial. UAVs can be obstructed from one another in a three-dimensional airspace \cite{ahmed2022recent}. Hidden nodes can cause transmission difficulties. Interference from other networks or ground-based systems can interfere with UAV network communications. Due to frequent channel sensing and re-transmissions, random access protocols may consume more energy. It is crucial to ensure secure communication, particularly when multiple UAVs are operating in sensitive environments. As UAV technology advances, the evolution of Random Access MAC protocols for UAV networks will continue. Future enhancements could include improved interference mitigation, advanced security mechanisms, energy-efficient protocols, and methods for administering complex three-dimensional airspace \cite{ahmed2022recent}.

\textbf{Time Division Multiple Access} (TDMA) is a MAC protocol widely used in various networking scenarios to manage communication between multiple devices efficiently. It works by dividing the available communication time into specific time slots or intervals, where each device is assigned its unique time slot for data transmission and reception \cite{kalwar2020hybrid}. In IoT-IoT networks, where numerous IoT devices need to communicate, TDMA brings structure and predictability to the communication process. Each IoT device is allocated a specific time slot to transmit or receive data. This time slot assignment reduces the likelihood of data collisions, ensuring efficient and organized communication. Moreover, IoT devices can synchronize their activities with the assigned time slots, which helps conserve power by minimizing the need for constant channel sensing, making it well-suited for low-power IoT devices with sporadic data transmission requirements \cite{siddavaatam2020novel}. In UAV-IoT networks, TDMA plays a critical role in coordinating communication between UAVs and the ground-based IoT devices they interact with. UAVs are assigned dedicated time slots for data collection, relaying information, or issuing commands. This allocation of specific slots ensures that UAVs can communicate with a large number of IoT sensors in a controlled and efficient manner \cite{vista2023hybrid}. The synchronization of time slots minimizes delays, essential for mission-critical applications like environmental monitoring, surveillance, and precision agriculture. TDMA also allows UAVs to operate efficiently by powering down during idle periods, thus conserving energy. In UAV-UAV networks, TDMA helps coordinate communication between multiple UAVs working together on missions such as search, rescue or surveillance. Each UAV is assigned a unique time slot for data transmission, preventing data collisions and interference. This ensures coordinated and efficient operations by UAVs in the network, essential for mission success and safety in complex airspace environments \cite{10217344}.

 \cite{8718117} proposes a dynamic time-division multiple access (DTDMA) scheme based on time mirroring to reduce relaying latency in airborne relay communication and to permit multiple nodes to access the network effectively. To accommodate a military communication environment in which nodes join and leave frequently, the proposed scheme also conducts efficient resource allocation and supports seamless voice communication. \cite{9349361} integrates an additional reservation period into a real-time, distributed, bidirectional TDMA scheme. This feature enables a node to reserve a future time slot while transmitting data in the present time slot. This mechanism not only reduces the control overhead associated with reserving time slots but also increases the network's efficiency by assuring maximum time slot utilisation. In addition, it improves spatial reuse, which is the practice of using the same frequencies for communications in various spatial areas, reusing the limited frequency spectrum. This is especially important in wireless networks with limited frequency resources. Collectively, these characteristics allow this TDMA scheme to provide greater throughput with very low control latency for both static and mobile network topologies.
%TDMA provides numerous benefits in a variety of network configurations. It provides predictable access to the communication medium, making it suitable for real-time or synchronised communication applications. By assigning unique time periods to devices or UAVs, TDMA reduces collisions and improves data transmission efficiency. IoT devices, UAVs, and other devices can conserve energy by shutting down during periods of inactivity, thereby improving their power efficiency. TDMA guarantees dependable and interference-free communication, which is essential for mission-critical applications.
Additionally, TDMA is scalable and can accommodate a large number of devices or UAVs through the addition of additional time segments \cite{shayo2020survey}. However, TDMA is also accompanied by obstacles and considerations. TDMA requires precise synchronisation among devices or UAVs for its efficacy. TDMA may not be the ideal option for applications with irregular or modest data transfer needs. Configuring and administering time slots for a large number of devices can be difficult and may necessitate centralization. Despite these obstacles, TDMA's advantages make it a valuable tool in various network scenarios \cite{shayo2020survey}.

\begin{figure}[!htb]
            \centering
            \includegraphics[width=.75\columnwidth]{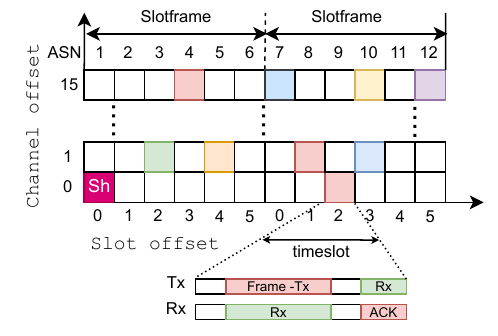}
            \caption{Scheduling of TSCH slots with slotframe size of $6$.}
            \label{fig:sf}
        \end{figure}

 \textbf{Scheduling Based Channel Access} is used in 6TiSCH kind of networks to transmit a sensory data packet from a given node to the border router or gateway. For this, the node must schedule a unique ``\textit{cell}'', which is a combination of a \textit{timeslot} (aka \textit{slot}) and a physical \textit{channel} (shown in Fig.~\ref{fig:sf}), with its' preferred routing parent. However, standards such as IEEE 802.15.4e or RFCs do not provide information about managing/scheduling (\textit{i.e.,} allocation and deallocation) such data transmitting cells. Scheduling of transmission cells is very challenging in dynamic IoT networks because assigning a larger number of cells would increase the energy consumption of the nodes by forcing them to activate their radios in the assigned slots. Conversely, fewer assigned cells would severely degrade the network performance. %and create network instability. 
 Therefore, a conflict- and collision-free cell scheduling scheme is expected to improve the performance of multi-hop  IoT networks.
 For scheduling cells in 6TiSCH networks, researchers used either centralized \cite{6362805}, distributed \cite{rfc9033,7273816}, or autonomous scheduling approaches such as Orchestra \cite{Duquennoy}. However, autonomous scheduling has the advantage of zero control packet overhead in the network and so can save nodes' energy. Recently, several autonomous schedulers have been proposed, such as Orchestra \cite{Duquennoy}, ALICE \cite{10.1145/3302506.3310394}, OST \cite{9155496}, and A3 \cite{kim2021a3}. However, both Orchestra and ALICE suffer from static and limited allocation of transmission (\texttt{Tx}) and reception (\texttt{Rx}) slots. On the other hand, OST and A3 provide adaptive slot allocation \textit{i.e.,} depending on networks' slot requirement, they suffer from high energy consumption when there is less traffic in the network. Additionally, OST and A3 allocate slots based on traffic estimation, which is not immediate. Consequently, it contributes to an increase in packet delivery latency. Atis \emph{et al.} extensively studied some of the existing autonomous scheduling methods using testbed experiments and suggested increasing the number of \texttt{Rx} slots at the root node in \cite{9032088}. However, increasing \texttt{Rx} slots of the root node only help the immediate neighbour nodes and significant performance improvement of the other nodes \textit{i.e.} two or more hop distance nodes cannot be expected. The works in \cite{9363324,10032650} autonomously allocate a shared cell for control packet transmission in 6TiSCH networks. The authors did not consider the transmission of data traffic in their works.

\color{black}
\subsection{Data Security and Privacy}

In this, we delve into the critical topics of data security and privacy, exploring two key areas: Physical Layer Security and Security in IoT systems. Regarding current communications and technological advancements, these domains are essential.
\subsubsection{\textbf{Physical Layer Security (PLS)}}

\begin{figure}[!t]
    \centering  \includegraphics[width=\columnwidth]{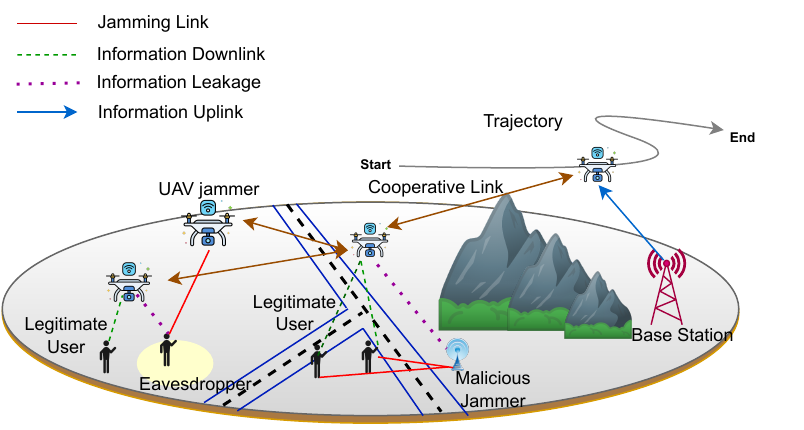}
    \caption{Physical Layer Security communication UAV network}    
    \label{fig_7:PLS}
\end{figure}

Traditional encryption methods require high computational complexity, which consumes a lot of energy. As an alternative, physical layer security makes use of the wireless channels' intrinsic uncertainty to secure wireless communication networks in a way that is computationally efficient and effective. In order to ensure communication security, several physical layer techniques have been presented in the past year. Typically as shown in Fig. ~\ref{fig_7:PLS}, a physical layer security communication network has three main nodes -- legitimate transmitter, legitimate receiver, and potential eavesdropper, which are modelled using a wiretap channel. Specifically, there are two forms of eavesdropping: passive eavesdropping intend to capture confidential data transmitted between legitimate link without degrading received signal quality at the receiver; on the other hand, in active eavesdropping, there is a degradation in main channel capacity by transmitting jamming signals to the receiver, at the same time also try to enhance the channel capacity of the eavesdropper. According to the role of UAV in communication networks, commonly considered PLS scenarios could be classified as follows:

\begin{itemize}
    \item \textbf{UAV as Transmitter or Receiver:} The UAV can transmit or receive data through the communication network. It actively shares information between authorized nodes, possibly leveraging its mobility and high altitude advantages to enhance security and communication quality.

    \item \textbf{UAV as Relay/Jammer:} As a relay, the UAV will transmit data from the authorized transmitter to the authorized receiver. On the other hand, it might potentially act as a jammer, aiming to improve the legitimate node's channel while weakening the communication link between the legitimate node and the eavesdropper.

    \item \textbf{UAV as Eavesdropper:} The UAV transforms into an eavesdropper, aiming to intercept sensitive data being passed between the legitimate transmitter and receiver. It employs ways to acquire this data discreetly and without discovery.

    \item \textbf{UAV as Hider or Surveillant:} The UAV can serve as a hider, trying to conceal legitimate communication from potential eavesdroppers. Alternatively, it can act as a surveillant, monitoring the network for any suspicious activity or eavesdropping attempts, thereby enhancing the overall security posture of the communication network.

\end{itemize}

In the past few years, various techniques were utilised to ensure physical layer security. In the domain of UAV systems, resource allocation takes centre stage, encompassing parameters like transmit power, cruising speed, time slot, and frequency bandwidth. It is crucial to acknowledge that this allocation doesn't just impact the signal strength for the intended recipient but also unwanted eavesdroppers. \cite{8644698} investigate the possibility of trajectory design to bolster physical layer security in UAV systems and introduce the concept of joint resource allocation and trajectory design. In this integrated approach, the UAV's flight path is carefully optimized to weaken eavesdropping links while strengthening legitimate ones. This optimization is obtained through iterative solutions to approximation problems. By considering a scenario with a UAV acting as a mobile relay, the goal is to maximize spectral efficiency while ensuring secure communication, even under practical constraints.
% \textcolor{red}{Which Approach??? I think you missed the refs for this para} 
This approach proves highly effective in enhancing physical layer security and spectral efficiency when the full deterministic position of the eavesdropper is known, which is not the case in real world. Additionally, a robust joint design strategy in \cite{8422581} is explored for scenarios with partial or statistical eavesdropper position information. Here, the UAV's trajectory optimizes the legitimate channel's capacity while avoiding uncertain eavesdropper areas. Furthermore, transmitting artificial noise to disrupt eavesdroppers and efficient power allocation for this purpose are discussed in \cite{wang2020artificial}, acknowledging the computational challenges of finding optimal power allocations in UAV-based systems. Thus, pursuing computationally efficient sub-optimal solutions remains a key concern in UAV-based wireless communications. Some advanced techniques can also be implemented into the UAV systems to further improve the system's secrecy performance. In the following section, we discuss how NOMA, beamforming, and Intelligent Reflecting Surface $($IRS$)$ technologies are used in UAV systems to enhance physical layer security.

\textbf{Secure Communication via Collaborative Beamforming}: Beamforming is a vital technology in wireless communications that optimizes signal direction to improve both security and efficiency \cite{wong2017key}. Traditional beamforming, operating in two dimensions (2D), effectively enhances physical layer security by directing signals away from potential eavesdroppers. When equipped with complete knowledge of Channel State Information (CSI), the UAV can steer beams orthogonal to eavesdropper channels, ensuring secure information transmission. However, this approach necessitates a trade-off between bolstering signal strength for intended users and degrading signal quality for eavesdroppers. In contrast, 3D beamforming, generating separate beams in a three-dimensional space, offers superior service coverage and throughput compared to its 2D counterpart \cite{zeng2018cellular}. This technique becomes particularly advantageous when users are distributed in a 3D space featuring varying elevation angles relative to the transmitter. Due to the high altitude of UAVs, intended receivers and potential eavesdroppers can be naturally distinguished by their distinct altitudes and elevation angles to the UAVs. The LoS channel characteristic in UAV systems enables precise beamforming both in azimuth and elevation domains. A narrow and precise beam can be shaped to efficiently transmit to desired legitimate receivers while minimizing information leakage possibilities to potential eavesdroppers.

\textbf{Secure Communication via NOMA}: NOMA is recognized as a highly promising technique for achieving superior spectral efficiency by multiplexing information signals at varying power levels \cite{wong2017key}. It is anticipated that implementing NOMA could significantly augment the achievable rate and resilience in UAV communications at the physical layer, focusing on security. In a given scenario where a UAV serves as a relay to facilitate data transmission to receivers with distinct security clearance levels within a maximum cruising duration T, the receiver with lower security clearance poses a potential eavesdropping threat, potentially intercepting signals intended for the receiver with higher security clearance. In instances where the eavesdropper experiences suboptimal channel conditions, NOMA is employed to transmit both confidential and public information concurrently. Conversely, if the eavesdropper's channel conditions are favourable, the UAV exclusively transmits public information for security reasons. The selection between NOMA and unicast modes is determined based on the outcomes of the proposed resource allocation optimization. Specifically, to maximize spectral efficiency, an integrated approach involving optimizing the transmission scheme, resource allocation, and UAV trajectory planning is crucial. \cite{wang2020uav} investigated a downlink secure UAV-NOMA network in which users are classified as security-required users and QoS-required users. %, with the QUs potentially acting as internal Eves.

\begin{figure}[!htb]
            \centering
            \includegraphics[width=.6\columnwidth]{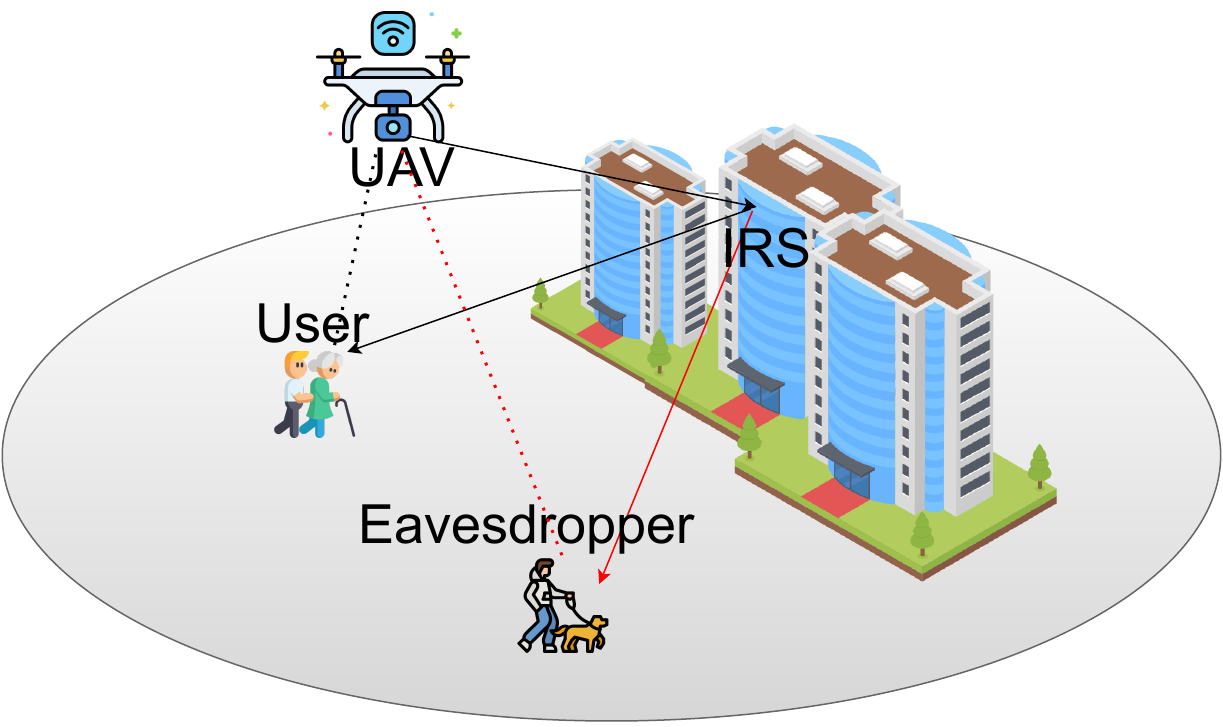}
            \caption{IRS aided UAV network}
            \label{fig:irs}
        \end{figure}

 \textbf{Secure Communication via IRS-assisted system}: The IRS is a revolutionary technology composed of specialized meta-surfaces designed to redirect incoming signals towards desired destinations, enhancing communication efficiency. Its adaptability, allowing precise control over phase shifts in its elements, enables optimization of signal reception at intended endpoints \cite{javed2022reliable}. IRS deployment options are versatile; one of them is shown in Fig. ~\ref{fig:irs}, including standalone setups or integration into buildings, especially in areas with inadequate signal coverage, thereby improving NLoS communication reliability. An intrinsic feature of IRS is its potential to significantly enhance UAV communication security and confidentiality by strategically modifying incident wireless signals, aiding in combating eavesdropping attempts \cite{wang2020intelligent}. Furthermore, the IRS offers the dual benefit of energy-efficient communication and a potential line of defence against security threats, ultimately extending UAV battery life. Through its application, IRS contributes to the enhanced PLS of UAV communications by constructively reinforcing signals received by the intended user and strategically interfering with unauthorized eavesdroppers. It finds applications in diverse UAV communication use cases, notably improving privacy and link security in scenarios such as mobile IRS-enhanced UAVs, NLoS scenarios, satellite communications, and cooperative jamming mitigation. 
In \cite{chu2020intelligent}, an algorithm fine-tunes UAV transmit power, location, and IRS phase shift to optimize secrecy rates in UAV-to-Ground User transmissions, outperforming baseline approaches. In \cite{long2020joint}, the focus is on maximizing network secure energy efficiency (EE), achieving this by optimizing user association, power control, trajectory, and phase shift, illustrating enhanced secure EE. \cite{li2021robust} considers UAV and ground user communication with IRS against a single Eve, optimizing UAV trajectory, IRS beamforming, and users' transmit power, highlighting IRS's positive impact on PLS and secrecy rates. Lastly, \cite{fang2021secure} introduces an iterative algorithm maximizing secrecy rates for BS-to-User communications, strategically controlling IRS phase shifts and UAV position, and utilizing the UAV as a passive relay for improved PLS.

\subsubsection{\textbf{Security in IoT}}
Security in IoT systems cannot be avoided and always presents a challenge. Due to the resource-constrained IoT devices, the outside attackers get more options to attack the IoT networks. Another critical aspect is that data flowing in a typical IoT system is not just ``data''; manipulation of such IoT data can severely damage in the system. For example, the door to our houses can be opened unnecessarily, or alarms can be turned on/off remotely. State-of-the-art security schemes are necessary to be ahead of these kinds of unwanted events. IoT networks such as IPv6 over Low-power Wireless Personal Area Networks $($6LoWPAN$)$ take advantage of the strong AES-128 link layer security defined in IEEE 802.15.4 standard. Basically, the link layer security provides link authentication and encryption to resource-constrained IoT devices. In addition to link layer security, transport layer security $($TLS$)$ mechanism is also very much needed in IoT networks. TLS, as defined in RFC 5246, runs over TCP. However, when UDP is chosen as the transport layer protocol, the RFC 6347 datagram transport layer security $($DTLS$)$ can provide security at the transport layer. However, it should be noted that implementing TLS/DTLS requires the device to have the necessary resources, such as a hardware encryption engine to enable the use of advanced cypher suites, etc. A device specially developed for this purpose is \textit{TI's CC2538 wireless MCU}, which integrates a powerful \textit{ARM Cortex-M3 CPU} and an \textit{IEEE 802.15.4 radio}. The device has up to 512kB Flash and 32kB RAM and a hardware encryption engine capable of supporting TLS/DTLS. The application layer protocols like CoAP (Constrained Application Protocol) can be secured with DTLS to encrypt and authenticate communication, providing security to IoT networks, whereas MQTT (Message Queuing Telemetry Transport) uses TLS for encryption and supports username/password authentication and access control lists (ACLs) to enhance the security of IoT data transmissions.

\color{black}

\section{Efficient data collection techniques in UAV-assisted IoT}
\label{sec:efficient}
% \subsection{Metrics for Assessing Reliability and Efficiency}
% \subsubsection{Energy Efficiency}
% \subsubsection{Time Efficiency}
% \subsubsection{Age of Information}
% \subsubsection{Throughput}
% \subsubsection{Data Rate}
% \subsubsection{Energy Consumption}
% \subsubsection{Outage Probability}
% \subsubsection{Average Bit Error Rate}
% \subsubsection{Cost Efficiency}
% \begin{itemize}
%     \item UAV deployment cost
% \end{itemize}

In this section, we discuss the key components of optimizing data collection in UAV-assisted IoT environments. We explore various techniques, such as trajectory planning and sensor network clustering, including UAV swarm coordination,  data aggregation and AI-based optimization. These strategies play an important part in optimising efficiency, reducing energy usage, and ensuring the uninterrupted functioning of IoT systems.

\subsection{Trajectory Planning}
Trajectory planning is subdivided into two essential components: path planning and collision avoidance, which are briefly discussed as follows.
\subsubsection{Path Planning}

Path planning in UAV flight is a critical task that involves determining the most efficient route from the UAV's starting point to its intended destination. Traditional path-planning methods often rely on known mathematical models of the environment, but this approach can be limiting, especially when operating in unknown or unpredictable environments. In the pursuit of efficient navigation through intricate surroundings, Wang et al. \cite{wang2019autonomous} explored deep reinforcement learning (DRL). DRL involves training neural networks to interpret sensory data and make informed decisions. In this context, the raw sensory data gathered by the UAV is converted into flight control signals, enabling autonomous and informed manoeuvring through complex environments. Collision avoidance is a crucial aspect of path planning to ensure the UAV's safety and prevent accidents with obstacles or other moving objects. Lin et al. \cite{lin2017sampling} developed a method that involves sampling-rooted path determination based on the swiftly exploring random tree algorithm. This algorithm effectively creates trajectories that avoid collisions with obstacles in real-time, considering factors such as the number of obstacles, their approach angles, and speeds. Pham et al. \cite{pham2018autonomous} introduced a novel approach employing reinforcement learning, a machine learning technique, to navigate UAVs effectively in uncertain and variable conditions. This technique involved training a UAV to reach its target using a PID+Q-learning algorithm, sidestepping the need for a precise mathematical model of the environment. The adaptability of this approach makes it particularly useful for real-world scenarios characterized by uncertainties.

Incorporating machine learning into path planning, Yijing et al. \cite{yijing2017q} proposed a strategy that synergistically combined Q learning and neural networks to enhance the UAV's learning rate during path planning and obstacle avoidance. By utilizing neural networks to model the UAV's continuous state space, the UAV could navigate and effectively escape challenging situations, thereby enhancing the efficiency and safety of the path computation process. The introduction of the Adaptive and Random Exploration (ARE) approach reduced learning errors, enabling more accurate computation of the route by allowing the UAV to adapt its actions based on real-time evaluations and redirect to a safe path when close to obstacles. Shiri et al. \cite{shiri2019massive} tackled the challenge of managing large UAVs, focusing on efficient movement and energy conservation, particularly in adverse conditions like windy weather. They employed a machine learning-assisted mean-field game approach, a mathematical framework that effectively guided UAVs to avoid mid-air collisions while minimizing communication and computational energy. Table~\ref{UAVtrajectory} for trajectory and path planning algorithms provides a comprehensive comparison of approaches that address autonomous UAV navigation and collision avoidance, offering insights into their strengths and limitations.

\subsubsection{Collision Avoidance}

Ensuring collision-free operations is a fundamental element of safe and efficient UAV flights. Systems like Sense and Avoid (SAA) or Detect and Avoid (DAA) play a pivotal role, enabling UAVs to seamlessly integrate into civil airspace and conduct operations Beyond Visual Line Of Sight (BVLOS) with a high degree of autonomy. Various methodologies exist to mitigate UAV collisions, such as utilizing GPS for obstacle detection and flight path adjustments, employing obstacle detection and avoidance sensors like LiDAR and sonar, leveraging computer vision through cameras to analyze and navigate the UAV's surroundings, utilizing AI-driven flight planning algorithms for intelligent obstacle avoidance and optimal flight planning, and promoting communication and coordination between UAVs and other aircraft to avoid potential collisions \cite{chhikara2020dcnn}.

The use of a monocular camera for UAVs to autonomously avoid collisions with obstacles in a variety of indoor scenarios was suggested by Singla et al. \cite{singla2019memory}. This method uses DRL and takes advantage of partial observability to allow UAVs to store important environmental information for sensible manoeuvring choices. The effectiveness and power efficiency of the method are greatly improved by including RNNs and deliberately reducing UAV oscillatory motion.  To quickly and precisely identify obstacles, Lai et al. \cite{lai2020detection} paired a monocular camera with a You Only Look Once (YOLO) algorithm and DRL. Joshi et al. \cite{joshi2023sim} looked into how measurement uncertainty affected UAV performance in waypoint navigation and obstacle avoidance using DRL. A Gaussian probability distribution is used to characterise this uncertainty, which results from sensor noise impacting the quality of obstacle recognition, and a Kalman filter is then used to refine the model further. Due to advancements in computer power, AI algorithms, and sensor technologies, the field of UAV collision avoidance is expanding quickly. These developments significantly improve UAV autonomy and safety, creating possibilities for wider deployment across different applications.

\begin{table}
\centering
\caption{Comparison of Trajectory and Path Planing Approach}
\resizebox{\columnwidth}{!}{
\tiny
\begin{tabular}{|p{0.1\columnwidth}|p{0.16\columnwidth}|p{0.16\columnwidth}|p{0.16\columnwidth}|p{0.14\columnwidth}|p{0.11\columnwidth}|}
\hline
\rowcolor[HTML]{8FEDDD}
\textbf{Articles} & \textbf{Path Planning Approach} & \textbf{Collision Avoidance Method} & \textbf{Strengths} & \textbf{Shortcomings} & \textbf{Use Case}  \\
\hline
Wang et al. (2019) \cite{wang2019autonomous} &
Deep Reinforcement Learning &
DRL for Collision Avoidance &
Autonomous navigation, collision avoidance &
Requires substantial computational resources &
In complex environments, autonomous navigation \\
\hline
Lin et al. (2017) \cite{lin2017sampling} &
Sampling-rooted path determination &
Swiftly-exploring Random Tree &
Real-time obstacle avoidance, adaptability &
Highly dynamic environments may pose challenges &
Real-time obstacle avoidance, dynamic scenarios  \\
\hline
Pham et al. (2018) \cite{pham2018autonomous} & Reinforcement Learning (PID+Q-learning) &
No precise mathematical model required &
Adaptability, effective in uncertain conditions &
Training and optimization may be time-consuming &
Real-world scenarios with uncertainties  \\
\hline
Yijing et al. (2017) \cite{yijing2017q} &
 Q Learning + Neural Networks &
Neural Networks for State Space &
Enhanced learning rate, obstacle avoidance &
Requires neural network model and training &
Learning rate improvement, obstacle avoidance  \\
\hline
Shiri et al. (2019) \cite{shiri2019massive} &
Mean-field Game Approach &
Mathematical Framework &
 Collision avoidance, energy conservation &
Complex mathematical framework&
Efficient movement, energy conservation  \\
\hline
\end{tabular}
}
\label{UAVtrajectory}
\end{table}

\subsection{Sensor Network Clustering}
Due to the limited energy of the UAV, the energy consumption caused by the flight of the UAV must be reduced in the scenario of UAV-assisted data collection. The framework of the sensor network becomes complex, and the cost of routing between sensors increases as the number of sensors increases. As a result, clustering must be used to separate the network into various sub-networks. The fundamental concept behind clustering is to combine nearby nodes into clusters and choose a CH from each cluster to be in charge of collecting data from other sensors in the cluster. Many clusters will raise the energy required for UAV flight but will minimize the energy needed for the cluster to forward data. In contrast, a few clusters will result in a reduction in the amount of energy used by UAVs during flight but an increase in the energy used by sensors. Thus, the energy consumption of sensors, the energy consumption of UAVs, sensor clustering, CH selection, and UAV path planning must all be concurrently optimized.

\begin{table*}
\centering
\caption{Summarise CH Selection Techniques}
\setlength{\extrarowheight}{2pt}
\resizebox{\textwidth}{!}{
\begin{tabular}{|l|p{0.16\textwidth}|p{0.16\textwidth}|p{0.16\textwidth}|p{0.16\textwidth}|p{0.11\textwidth}|p{0.16\textwidth}|p{0.16\textwidth}|p{0.16\textwidth}|}
\hline
\rowcolor[HTML]{8FEDDD}
\textbf{Algorithm Name} & \textbf{CH Selection Algorithm} & \textbf{Energy Optimization Approach} & \textbf{Sensor Clustering Method} & \textbf{CH Selection Criteria} & \textbf{Strengths} & \textbf{Shortcomings} & \textbf{Use Case} \\
\hline
Ghorbel et al. (2019) \cite{ghorbel2019joint}& Multi-hop clustering algorithm & Decreasing number of hops & Hierarchical clustering & Shortening flight times & Reduced flight times, energy-efficient data forwarding & Potential for limited coverage area & When flight time reduction is a priority \\
\hline
Tong et al. (2019) \cite{tong2019uav} & Affinity propagation algorithm & Joint CH and UAV optimization & Cluster-based CH selection & Reduction of Age of Information & Improved CH selection and flight path & May not consider sensor energy constraints & When reducing Age of Information is crucial \\
\hline
Albu et al. (2018) \cite{albu2018energy} & Maximum sensor power as CH & Value of Information (VoI) & Distributed clustering & Maximizing sensor lifespan & Energy-efficient, VoI-based data forwarding & Potential energy inefficiency & When maximizing sensor lifespan is a priority \\
\hline
Chen et al. (2019) \cite{chen2019efficient}& Maximum sensor power as CH & Value of Information (VoI) & Distributed clustering & Maximizing sensor lifespan & Reduced energy consumption by sensors & Potential to delay data forwarding & When reducing sensor energy consumption is essential \\
\hline
\end{tabular}
}
\label{CH}
\end{table*}

\subsubsection{Cluster Head Selection}

CH selection is a critical task in clustering. It involves selecting a single node or a subset of nodes inside a cluster to serve as its coordinator.  Certain criteria, such as the node's energy level or communication capability, or a combination of these aspects, influence this choice \cite{6720617}. To save time and energy, \cite{ghorbel2019joint} considers multi-hop clustering, which modifies the number of hops in each cluster, and CH is selected so that overall flight time minimises. The small number of clusters is determined by a larger cluster, which shortens flight times and narrows the UAV's path. On the other hand, a small cluster lowers the energy requirement for data forwarding in the sensor network. \cite{tong2019uav} collaboratively refined the CH selection and flight path of the UAV to reduce the AoI, using an algorithm based on affinity propagation to locate the CH and the related sensor. The energy restrictions on sensor clustering were considered in \cite{albu2018energy} and \cite{chen2019efficient}. To increase the lifespan of sensors, \cite{chen2019efficient} chose the sensor with the maximum power as the CH. The sensors in the cluster will only send data to the CH when the Value of Information (VoI), which is highest when an event first occurs and decreases over time, is greater than a predetermined threshold. This significantly reduces the energy used by the sensors. To reduce the average sending distance from the cluster's sensors to the CH, \cite{albu2018energy} used a distributed clustering approach. Table ~\ref{CH} summarises these techniques.

% \textcolor{red}{What is the difference between these two sections Cluster Head Selection and CH Selection Algorithm? Shouldn't be in the single section??}

\subsubsection{Clustering Algorithm}

The clustering algorithm is a mechanism for grouping nodes or devices inside a network into clusters. Its primary objective is to create logical groups of nodes to improve network efficiency, data processing, and communication while also supervising the management of network operations. It assigns nodes to specific clusters according to predefined criteria or rules. These criteria may include node proximity, energy levels, communication quality, and other network-specific indicators. K-means is a common clustering method in UAV networks due to its effectiveness and simplicity. It rapidly clusters the whole network, which is particularly beneficial in dynamic applications where network topology changes often. Various K-means-based techniques are discussed for efficient clustering in \cite{suma2021evolutionary}. A study \cite{he2015robust} presented the constrained coverage $($CC$)$ strategy, which took into account K-neighbors for each cluster by utilizing two virtual forces. However, this technique may result in a reduction in the sensor network's lifetime and a limited coverage area of the network. The Low-energy adaptive clustering hierarchy $($LEACH$)$ is a classical protocol that introduces the concept of clustering within a WSN and introduces a hierarchical approach to data transmission \cite{heinzelman2000energy}. This clustering technique organizes the WSN into groups or a hierarchy of clusters, which collect data from their surroundings and transmit it to their respective CH. The strategic selection of CHs within a WSN cluster can maximize the communication range and extend the network's lifespan. In each round, the method randomly designates CHs in a stochastic manner. Subsequently, the selected CH communicates with every non-CH node in the cluster to gather sensed data. The selection of the optimal CH is a crucial task, as several criteria must be met to identify the best node within the entire cluster \cite{chatterjee2000demand}. These criteria encompass factors like remaining energy, range, throughput, and mobility of each sensor node.

Although the LEACH algorithm prolongs the network's lifespan compared to multi-hop and direct transmission, it has several limitations. Selection is based on random choices, which do not guarantee an optimal solution and can result in an uneven distribution of SNs within each cluster, leading to imbalance. Nodes with lower residual energy levels are given the same priority as those with higher residual energy levels for CH selection. Consequently, when an SN with lower energy is chosen as a CH, its energy depletes more rapidly, reducing the network's overall lifetime \cite{khan2021noma}. In \cite{9619857}, enhanced research was presented that leverages the LEACH algorithm to enhance the energy efficiency of WSNs. Several alternate strategies have also been presented. For instance, \cite{loscri2005two} introduced the TL-LEACH method, a two-tier hierarchical approach that efficiently distributes the energy load across Sensor Nodes (SNs) in vast networks. \cite{4394931} developed an energy-LEACH (E-LEACH) technique by incorporating SNs' leftover energy into LEACH during the CH selection procedure. E-LEACH does not consider the distance between the Base Station (BS) and CHs, instead focusing on reducing the average distance between CHs and non-CH nodes. A voice-LEACH (V-LEACH) protocol was developed by Yassein et al.\cite{bani2009improvement} that names a voice-CH within the cluster in addition to choosing a CH. When the primary CH is exhausted, the voice CH takes over CH's responsibilities. Although V-LEACH is more energy efficient than LEACH, it still uses more energy when choosing voice-CHs for clustering.

The game theory-based energy-efficient clustering (GEEC) routing protocol was recently proposed by Lin et al. \cite{lin2017game} by applying a game model to CH selection. To establish energy balance and extend network longevity, GEEC uses a mechanism based on evolutionary game theory. The energy-efficient compressive sensing-based clustering routing (EECSR) protocol was created by Wang et al.\cite{wang2019energy} and combines the advantages of clustering techniques and compressive sensing-based schemes. The PSO-C algorithm for energy-efficient clustering in WSNs was introduced by Latiff et al. \cite{latiff2007energy} PSO-C takes into account variables like the average intra-cluster distance, the sum of all SN beginning energies, and the sum of all CH current energies. However, PSO-C may decrease network energy efficiency by assigning non-CH nodes to the closest CH during cluster formation. Additionally, Singh et al. \cite{singh2012novel} introduced a PSO-based energy-efficient CH selection technique known as PSO-semi distributed (PSO-SD), applying a fitness function that takes variables like residual energy, distance, and node density into account. However, the cluster formation stage, which can result in a decline in network energy efficiency, is not addressed by PSO-SD. In \cite{srivastava2015genetic}, an optimized zone-based energy-efficient protocol $($OZEEP$)$ was proposed to optimize CH selection and enhance clustering by incorporating genetic fuzzy systems $($GFS$)$. Optimizing CH selection and improving the cluster structure are critical challenges in clustering.  

\subsection{Data Aggregation}
% \textcolor{red}{Rewrite the below sentence. At first, write something about what is data aggregation.}
% \textcolor{blue}{It is an effective and precise method of gathering data in remote locations, UAV-based data aggregation addresses issues such rapid mobility, dynamic topological changes, and sporadic communication links \cite{arafat2020routing}.}

Data Aggregation refers to collecting and analysing data from various sensors in order to reduce redundant transmission. This procedure often involves combining data from several sensors at intermediate nodes and transferring the resulting aggregated data to a base station. This strategy is very beneficial when there are too many data points or the network architecture is changing. Data aggregation can help improve network energy efficiency, enhance the lifespan of wireless sensor networks, and minimise network traffic by minimising the quantity of data that needs to be transmitted \cite{abbasian2020survey}. 
Data aggregation in a UAV-assisted network begins when sensors start delivering data towards the aggregation sites i.e., CH, and concludes when the UAV reaches the last aggregation point on its route, here UAVs serves as ariel base station \cite{9377438}. 

To improve the effectiveness of data collecting, Wang et al. \cite{wang2019uav} developed the TA-UAV-DA protocol, which was motivated by compressive sensing. They constructed a balanced tree-based topology, which reduces the amount of measurement and updated data. Better data reconstruction efficiency is achieved when Centralized Nodes (CNs) broadcast data to CHs, and UAVs collect data from these CHs. The EE-UAV-DA protocol was developed by Wu et al. \cite{wu2018energy} and uses UAVs as data mules. This protocol uses a genetic algorithm to optimize data mule routes, resulting in great energy efficiency while minimizing latency between sensor nodes and the sink node. A heuristic search technique effectively reduces energy usage by optimizing CH selection and data mule routes. To improve communication and processing for multi-UAV information collection, \cite{thammawichai2017optimizing} proposed OC-mUAV. To reduce energy consumption, multi-hop clustering and adaptive energy consumption models are used. With multi-hop networks and the clustering of different types of UAVs, this self-organized network increases flexibility, dependability, and network longevity. A data-gathering algorithm using UAVs and mobile agents (MAs) for disaster site reconnaissance was introduced by Dong et al. \cite{dong2014uav}. Data gathering from sensor nodes and communication with UAVs are the responsibilities of MAs. Information-driven static and dynamic MA planning algorithms enable efficient routing, making it appropriate for crowded networks. Liu and Zhu \cite{liu2019energy} optimized data transmission techniques to address energy efficiency in UAV-aided WSNs. Sensor nodes make dynamic transmission mode selections to save energy and guarantee timely data delivery.

The performance of an energy-constrained IoT system using a power beacon and UAVs for data collection is examined in \cite{cvetkovic2023capacity}, offering light on the viability and limitations of IoT devices, including UAVs in situations with restricted energy resources. The construction of a fully immersive industrial metaverse is explored in \cite{li2022completion}, with a focus on the optimization of key performance indicators such as reliability, latency, and the Age of Information using creative packet architectures in 6G URLLC communication. \cite{nie2023novel}, on the other hand, introduces a cooperative strategy involving unmanned ground vehicles (UGVs) and UAVs for data collection from sensor nodes. This method efficiently optimizes data collection in scenarios where both ground and aerial vehicles cooperate to overcome obstacles caused by sensor node limitations.

\subsection{UAV Swarm Formations}

Using multiple UAVs or drones, working together is more effective for challenging tasks compared to a single UAV. This collaborative approach, known as Flying Ad-Hoc Networks (FANETs), faces significant challenges like fast mobility, frequent topology changes, unreliable connectivity, and limited energy. These issues impact operational reliability and efficiency. Clustering UAVs for hierarchical management is an effective strategy for addressing these problems and improving performance. Clustering aids in the control of node heterogeneity, as well as the improvement of fault tolerance and scalability. Each UAV becomes a node in this system, and nodes are clustered into clusters. Each cluster has a CH who is in charge of handling communication both inside and between clusters, and the other UAVs in the cluster are known as Cluster Members (CMs). Effective CH selection and management lower communication overhead, improve network performance, and allow for direct communication within the effective range. The existing algorithms used in UAV swarm clustering are categorised into four groups.

% \textcolor{red}{What are the below subsections?? Schemes for swarm formation ?? Looks like suddenly they have appeared. Also no reference like how the existing works have done swarm formation. We need to discuss the works in Table VII}

\subsubsection{Considering Localisation}

A data routing scheme that combines localization and energy efficiency is introduced in one technique in the literature \cite{khelifi2018localization}. To improve localization accuracy, this method makes use of signal strength data, weighted centroid localization, and fuzzy logic. In order to reduce total energy consumption and lengthen cluster lifetimes, it also optimizes the choice of CHs, taking distances and remaining energy into account. Furthermore, using particle swarm optimization (PSO),\cite{arafat2019localization} propose Swarm Intelligent Localization (SIL). Using anchor node distances, SIL sharpens the search space, accelerates convergence, and more precisely calculates UAV positions. Additionally, they offer a Swarm Intelligent Clustering (SIC) algorithm that lowers energy use by taking position information, distances between and within clusters, and residual energy into account. Additionally,\cite{arafat2021bio} offer Bio-inspired Localization (BIL) and Bio-inspired Clustering (BIC) algorithms that optimize cluster formation and head selection while cutting computing costs and effectively reducing energy consumption. These clustered routing techniques with location-based localization not only provide improved localization accuracy and energy efficiency but also contribute to superior Packet Delivery Ratio $($PDR$)$. They support more reliable and effective UAV networks by being well-suited for circumstances requiring strict localisation, such as post-disaster search and rescue missions.

\subsubsection{Considering Mobility}

Mobility-considered algorithms are adapted to dynamic network conditions, optimizing routing and communication while minimizing delays and ensuring reliable data delivery. The frequent shifts in network topology due to UAV mobility have prompted the introduction of a topology-aware routing method known as TARU for UAV swarms \cite{hong2020toward}. TARU begins by calculating link durations based on formation status and topology-related data, dynamically adjusting parameters like hello intervals and link hold times. This optimization reduces average delays and enhances routing performance. However, in cases of rapidly changing neighbour relationships, the hello message overhead increases, calling for potential improvements. \cite{shu2011mobility}, propose Mobility Predictive Clustering Routing $($MPCR$)$, selecting CHs based on node connectivity and introducing ferry nodes to bridge disconnected clusters. MPCR excels in terms of delay and data delivery but does not account for node energy levels, limiting its use for UAVs with limited battery capacity. \cite{6162360}, introduce Mobility Prediction Clustering Algorithm $($MPCA$)$, tailored to UAV characteristics. MPCA employs a Link Expiration Time $($LET$)$ mobility prediction model and a Dictionary Trie Structure $($DTS$)$ prediction algorithm for speedy cluster updates. CH selection relies on a combination of LET and DTS predictions, making it suitable for highly dynamic FANETs. Furthermore, the literature presents Mobility and Location-aware Stable Clustering $($MLSC$)$ \cite{9109267}, designed to enhance the performance and reliability of resource-constrained UAV networks. MLSC optimizes cluster size, employs distance-based k-means clustering for CH selection, and constructs a backbone routing tree using an approximate Minimum Spanning Tree $($MST$)$. It introduces an active backup CH scheme for CH location maintenance, reducing network overhead, latency and enhancing packet delivery rate. However, it does not consider node energy levels, suggesting potential energy-based optimizations. 

In summary, these clustering routing protocols address challenges posed by UAV mobility and frequent topology changes. They prioritize factors like node connectivity and link duration, delivering good PDR and End-To-End Delay $($ETED$)$ performance, suitable for time-sensitive scenarios like rescue missions.

\begin{table*}
\centering
\caption{Comparison of UAV-Swarm Formations Schemes}
\resizebox{\textwidth}{!}{
\tiny
\begin{tabular}{|p{0.1\textwidth}|p{0.12\textwidth}|p{0.08\textwidth}|p{0.11\textwidth}|p{0.14\textwidth}|p{0.16\textwidth}|}
\hline
\rowcolor[HTML]{8FEDDD}
\textbf{Articles} & \textbf{Localization} & \textbf{Topology-Aware} & \textbf{Efficiency} & \textbf{Strengths} & \textbf{Shortcomings}  \\
\hline
Khelifi et al. (2018) \cite{khelifi2018localization} & Signal Strength, Centroid & No & Energy Optimization & Improved Localization, Energy Efficiency & Not Suitable for Highly Mobile UAVs \\
\hline
Arafat et al. (2019) \cite{arafat2019localization} & Anchor Node Distances & Yes & Energy Efficiency & Improved Localization, Reduced Energy Consumption & Requires Computational Resources  \\
\hline
Arafat et al. (2021) \cite{arafat2021bio} & Bio-inspired Optimization & Yes & Energy Efficiency & Optimized Clustering & Reduced Energy Consumption  \\
\hline
Hong et al. (2020) \cite{hong2020toward} & No & Yes & Enhanced Routing Performance & Reduced Delays & Increased Hello Message Overhead  \\
\hline
Shu et al. (2011) \cite{shu2011mobility} & No & No & Delay and Data Delivery & Bridge Disconnected Clusters & Limited Battery Capacity  \\
\hline
Arafat et al. (2020) \cite{6162360} & Yes & Yes & Cluster Updates & Highly Dynamic FANETs & Highly Dynamic FANETs  \\
\hline
Raza et al. (2018) \cite{aadil2018energy} & No & No & Energy Optimization & Longer Cluster Lifetimes & Limited Support for Highly Mobile UAVs \\
\hline
Khan et al. (2019) \cite{khan2019bicsf} & No & Yes & Hybrid Approach & Extended Cluster Lives & Possible Overly Selective CHs \\
\hline
Fang et al. (2019) \cite{fang2019energy} & No & Yes & Energy Balance & Improved CH Selection & Enhanced Cluster Stability  \\
\hline
\end{tabular}
}
\label{UAVswarm}
\end{table*}

\subsubsection{Considering Energy Consumption}

Energy-efficient UAV swarm formation algorithms focus on reducing energy consumption to prolong the cluster's lifetime and improve data collection efficiency. They take into account the limited energy resources of UAVs to ensure reliable communication. Addressing these challenges, researchers have proposed innovative clustering-based routing protocols. Two notable models, the Energy Aware Link-based Clustering $($EALC$)$ \cite{aadil2018energy} and the bio-inspired clustering scheme for FANETs \cite{khan2019bicsf}, focus on enhancing data collection efficiency and ensuring reliable communication. The energy-efficient CH selection by a K-means density clustering technique is prioritized by the EALC model, which was first developed by Raza, A. et al.\cite{aadil2018energy} By taking into account node energy levels and closeness to neighbors, this method results in longer cluster lifetimes and less overhead. EALC includes a cluster maintenance method for additional optimization, albeit it might need to be improved to support highly mobile UAVs. The bio-inspired clustering BICSF \cite{khan2019bicsf}, technique reportedly uses a hybrid approach combining glowworm swarm optimization $($GSO$)$ and krill herd $($KH$)$ to overcome challenges brought on by energy limits and UAV mobility. BICSF effectively maintains clusters, improves inter-cluster communication, and ensures cluster stability. It also optimizes CH selection based on energy concerns. BICSF reduces energy usage while prolonging cluster lives thanks to its quick convergence to optimal solutions. However, it could also need to handle the possible problem of overly selective CHs. These clustering-based routing methods highlight the effectiveness of data collection and the dependability of communication by taking UAV energy levels into account and encouraging extended CH survival. These characteristics make them well-suited for situations with less mobility, ensuring reliable and efficient data transfer, making them perfect for jobs like reconnaissance and surveillance.

\subsubsection{Jointly considering Energy Consumption and Mobility}

Some algorithms consider energy consumption and mobility, offering a balanced approach that ensures reliable communication while optimizing energy usage. These algorithms often employ novel strategies to manage energy and adapt to dynamic mobility patterns. The Energy Balance and Mobility Prediction (EBMP) clustering approach was introduced in \cite{fang2019energy}. In the cluster creation phase, EBMP uses a novel model to calculate UAV energy consumption and stability by allocating weights to each UAV. This guarantees that the best UAV is chosen to serve as the CH. To improve cluster stability during cluster maintenance, EBMP uses a CH rotation approach. Inter-cluster communication uses gateways selected using the shortest distance projection method, whereas intra-cluster communication is direct. EBMP performs better in terms of average CH lifetime, throughput, and latency than MPCA and MPCR algorithms. In \cite{khan2021intelligent}, an Intelligent Cluster Routing Scheme (CRSF) was introduced for FANETs to address UAV mobility-related communication challenges. CRSF dynamically selects CHs based on UAV position and remaining energy using a Moth Flame Optimization (MFO) approach, improving cluster stability. The routing algorithm in CRSF factors in Euclidean distance and remaining energy for effective load balancing and routing in FANET. These clustered routing methods optimize network energy usage, throughput, and latency while extending cluster lifetimes by considering energy and node degree. They are well-suited for scenarios requiring low-latency communication and prolonged operation.

For various issues, including reducing the energy consumption of UAVs, preserving the stability of the cluster structure, and getting precise location data for UAVs, various protocols for clustering routing methods are given. The protocols are summarized and compared in Table ~\ref{UAVswarm} in terms of their underlying algorithms, the rationale behind CH selection, cluster maintenance techniques and strengths and shortcomings. 
%\section{Existing Optimized Solutions for Reliable and Efficient Data Collection}
\subsection{AI Optimisation Techniques}

In the rapidly evolving field of UAV systems, the integration of AI techniques has spurred numerous advancements, enabling UAVs to perform a wide array of tasks with increased efficiency and accuracy. We briefly discuss these AI techniques in greater detail, their appropriate use cases, and the advantages they offer over classical ML algorithms.

\subsubsection{Reinforcement Learning (RL)}

% Reinforcement Learning (RL) has played a pivotal role in enabling UAVs to make optimal decisions in dynamic and complex environments. Notable aspects of RL in UAV systems include its capacity for altitude control, path planning, autonomous flight control, and resource allocation \cite{}. RL algorithms, such as Deep Deterministic Policy Gradient (DDPG) and Proximal Policy Optimization (PPO), have been used to improve altitude control in UAVs, surpassing the limitations associated with traditional Proportional-Integral-Derivative (PID) control methods. This advancement is vital for ensuring precise altitude regulation in UAV operations \cite{}. Furthermore, RL is particularly well-suited for tasks demanding real-time, sequential decision-making. UAVs benefit from RL algorithms when it comes to path planning, autonomous flight control, and resource allocation. For instance, RL technology has been applied to optimize UAV paths and adapt to varying environmental conditions, ensuring that UAVs navigate effectively and efficiently through dynamic and complex scenarios \cite{}.

RL is a dynamic approach fundamental to autonomous decision-making in UAV systems. It enables UAVs to learn by trial and error, adapting to their environments to maximize rewards. At each interaction, the UAV processes sensory input, makes informed decisions, and accumulates rewards. The core idea of RL is to continuously improve decision-making over time by seeking to maximize cumulative rewards through learning and adaptation \cite{kaelbling1995introduction}. RL encompasses various approaches, including Model-Free, Model-Based, and Hybrid RL, each offering distinct strategies for learning and decision-making in dynamic environments.

  \textbf{Model-Free RL Algorithms:} In the UAV domain, Model-Free RL (i.e., Navigating Without a Map) algorithms stand out as a remarkable arsenal. These algorithms adapt their policies based on the outcomes of their actions without the need to explicitly model the intricate dynamics of the environment. Among them, there are two distinct subcategories : 
     \begin{enumerate}
     \item Policy-based algorithms: Policy-based algorithms focus on learning a policy that maps states to actions. This approach sidesteps the need to estimate value functions or action-value functions (Q-functions). Instead, they leverage gradient ascent algorithms to update parameters in the direction of an objective function's gradient, anticipating future returns \cite{cui2019multi}. The real-world implications of policy-based RL techniques are profound, as they find applications in revolutionizing UAV flight control systems. For instance, state-of-the-art RL algorithms, such as Deep Deterministic Policy Gradient and Proximal Policy Optimization, have been deployed to enhance the accuracy and efficacy of altitude control in UAVs \cite{koch2019reinforcement}.
    \item Value-based algorithms: Value-based algorithms base their decisions on estimates of value functions to maximize the anticipated cumulative payoff. This approach entails continual updates using techniques like temporal difference learning, which calculates the difference between actual and predicted values. In a UAV context, this method has been employed to address complex, real-world issues, such as long-term resource allocation. UAVs, acting as learning agents, execute resource allocation solutions via Q-learning-based frameworks to maximize rewards \cite{cui2019multi}.
    \end{enumerate}
    \textbf{Model-Based RL Algorithms:} Model-Based RL (i.e., Learning from the Environment) algorithms offer another dimension to UAV decision-making. These algorithms blend learned models of the environment with RL techniques, enabling UAVs to simulate possible actions and their outcomes for a better exploration and optimal policy determination \cite{qu2020novel}. Two main subcategories, imitation learning and inverse reinforcement learning, emerged as pioneers in this field. 
    \begin{enumerate}
     \item Imitation learning: It is often employed in a robotics context and rapidly suggests solutions to intricate control problems. Whether performed by humans or machines, imitation learning relies on leveraging a model of the system's dynamics. In UAV applications, this approach has significantly enhanced tracking performance, allowing UAVs to adapt to situations where adhering to a predefined trajectory might be impractical or counterproductive to system dynamics \cite{liang2018scalable}. Furthermore, this technique has been extended to multi-agent scenarios, offering insights into optimizing UAV deployment strategies that guarantee optimal user benefits and UAV owner profits \cite{wang2021dynamic}.
     \item Inverse reinforcement learning (IRL): It takes on the challenge of deducing an agent's reward function from observed behaviour. This method is valuable for tracking UAV paths and minimizing tracking errors. Expert demonstrations provide the foundational data, which is processed through hidden Markov models (HMM) and dynamic time warping (DTW) to generate representative trajectories. IRL then determines the hidden reward function, which guides the UAV in trajectory tracking with minimal error \cite{choi2017inverse}.
    \end{enumerate}
    \textbf{Hybrid RL Algorithms:} The fusion of multiple learning models defines Hybrid RL Algorithms, a dynamic approach to decision-making. In the UAV realm, model-ensemble-based learning finds prominence. This approach combines various learning models, such as Logistic Regression and Naive Bayes classifiers, to make predictions based on diverse data sources. This technique has found application in predicting critical attributes like crop yield or water quality \cite{qu2020novel}. One noteworthy application involves predicting wheat production using hyperspectral data and ensemble-based learning, demonstrating how machine learning techniques can have a substantial impact on agriculture \cite{li2022uav}. In the UAV patrolling context, where coordination among multiple UAVs is vital, a hybrid approach combines reinforcement learning with multi-agent simulations. This strategy enables UAVs to autonomously discover optimal patrolling patterns in real-world, geo-referenced environments, even when faced with unknown obstacles and dynamic targets. Coordination Learning in Multi-Agent Framework facilitates the development of this groundbreaking hybrid strategy \cite{perron2008hybrid}.

\subsubsection{Generative Adversarial Networks (GANs)}

It is a class of ML frameworks designed to generate new, synthetic instances of data that can pass for real, original instances. It consists of two parts: a generator that learns to generate convincing data and a discriminator that learns to discern between the generator's phoney data and actual data. The generator's task is to mislead the discriminator rather than to minimise the distance to a specific image. As a result, the model is able to learn unsupervised. GAN is currently popular in the field of UAVs. It is used for data generation and widely deployed for various purposes in the UAV context. One primary use of GAN is data augmentation in UAV systems. They generate synthetic data samples that closely mimic real-world scenarios, proving invaluable when actual data collection is restricted, expensive, or unfeasible, ultimately boosting the resilience of machine learning models trained on UAV data \cite{goodfellow2014generative} In addition, GANs, particularly Deep Convolutional GANs (DCGANs), play a significant role in UAV applications related to image and video production \cite{radford2015unsupervised, alladi2021drone}. DCGANs utilize convolutional neural networks (CNNs) as both the generator and discriminator, excelling in creating high-quality images from random noise. This technology greatly enhances the visual data generation capabilities of UAV systems. Furthermore, GANs address common challenges in UAV applications, such as data imbalances. For instance, in UAV-based wildfire detection, GANs like CycleGANs are pivotal in generating synthetic wildfire images. This effectively balances the training dataset and enhances the accuracy of wildfire detection models through synthetic wildfire image generation, contributing to the overall performance of UAV-based detection systems \cite{park2020wildfire}. Finally, GANs are also employed in object recognition tasks, assisting UAVs in identifying and categorizing objects within their field of view. These applications highlight the versatility and importance of GANs in UAV systems. Traditional ML algorithms are more versatile and easier to train but may not perform as effectively in scenarios where data generation and augmentation are critical. The choice between GANs and traditional ML algorithms depends on the specific requirements and challenges of the UAV application.

\subsubsection{Recurrent Neural Networks (RNNs)}
These are a type of artificial neural network designed to handle sequential data. Unlike classic neural networks, where all inputs and outputs are independent of one another, RNNs can manage sequences of inputs using their internal state stated as memory. As a result, these are especially beneficial for applications like time series prediction, natural language processing, and speech recognition \cite{9044359}. RNNs, especially with variants like Long Short-Term Memory (LSTM) and Gated Recurrent Unit (GRU) networks, are highly adaptable in UAV systems, given their exceptional capacity for handling sequential data. One of the notable strengths of RNNs is their ability to process temporal dependencies and manage sequential data effectively \cite{graves2013speech, heaton2018ian}. In the context of UAVs, LSTM networks have been employed in a multitude of applications, including resource allocation in Machine-to-Machine (M2M) communications \cite{xu2021generative}. LSTM networks, when combined with GANs, address resource allocation challenges in UAV networks, particularly in scenarios involving multiple UAVs. The synergy between LSTM and GANs has led to optimized UAV mobility and improved network performance, enabling UAVs to perform efficiently in complex and dynamic environments. Moreover, LSTM networks are proficient in anomaly detection. This is particularly valuable in scenarios where anomaly detection is vital for UAV system security and performance. By combining LSTM with Auto Encoders (LSTM-AE), UAV systems enhance stability through the detection of anomalies and irregularities, ensuring that they operate reliably in changing and challenging conditions \cite{bae2020uav}. Another application of RNNs, including GRU networks, is heading error modelling. In UAV systems, GRU networks are employed for heading error modelling to enhance the accuracy of forecasting direction and improve defect detection. For instance, these networks have been utilized to predict heading errors caused by body axis tilt during flight, further ensuring the precision and reliability of UAV operations \cite{zhao2022attitude}.

\subsubsection{Convolutional Neural Networks (CNNs)}

CNNs serve as the fundamental of image analysis in UAV applications. It extracts crucial features from images, enabling UAVs to identify objects, classify images, and detect anomalies \cite{chriki2021deep}. In the context of UAV-based surveillance and infrastructure inspection, CNNs automatically identify damage and anomalies, such as defects in wind turbine blades. They significantly improve the accuracy of damage detection and reduce the reliance on manual feature engineering. Through the application of CNNs, research enhances the performance of UAV systems by enabling the automatic identification of damage and anomalies, crucial for ensuring the structural integrity of critical infrastructure. The reduction in the need for manual feature engineering further streamlines the UAV's ability to perform precise damage assessments.

\subsubsection{Comparison and Application}

In summary, AI techniques such as GANs, RNNs, CNNs, and RL serve unique yet interconnected purposes within UAV systems. GANs primarily focus on data augmentation, image synthesis, addressing data imbalances, enhancing object recognition, surveillance, and image generation. RNNs handle temporal dependencies, resource allocation, anomaly detection, and predictive tasks, enabling better trajectory prediction and real-time decision-making. CNNs excel at image analysis, enhancing object recognition, damage detection, and anomaly identification, thus reducing the need for manual feature engineering. RL empowers UAVs to make optimal decisions in dynamic, real-time environments, benefiting tasks such as path planning, autonomous flight control, and resource allocation.

The choice among these techniques depends on the specific UAV application's requirements, whether it involves data generation, sequential data processing, image analysis, or dynamic decision-making. Each technique plays a crucial role in advancing UAV systems, pushing the boundaries of what these autonomous vehicles can achieve in various domains, from surveillance and image analysis to resource optimization and navigation. These AI techniques contribute to the increasing autonomy and capability of UAVs, allowing them to perform a wide array of tasks with efficiency and precision in dynamic and complex environments. The seamless integration of these AI technologies is paramount in realizing the full potential of UAVs across applications like surveillance, infrastructure inspection, and resource allocation.

\section{Applications and Use Cases}
\label{sec:applications}
% \subsection{Real-World Applications}
% \subsection{Use Cases in Various Domains}
% \subsection{Impact and Benefits}

In this section, we briefly discuss two use cases of UAVs to improve the reliability and efficiency of IoT networks.

\subsection{UAV as Mobile Edge Computing Server}

Compute-intensive applications such as autonomous driving and traffic control have significantly enhanced mobile user experience. However, ground devices (GDs) face challenges due to limited computing power and energy \cite{zhou2020mobile}. To tackle this, MEC has emerged as an advanced solution, shifting mobile computing to network edge nodes (e.g., base stations, access points), allowing compute-intensive applications on GDs with constrained resources \cite{mao2017survey}. Simultaneously, the widespread use of IoT devices has brought about increased convenience. Leveraging IoT with UAVs optimizes A2G and LoS transmission, ensuring reliable data transmission, especially in remote and high-traffic areas, while overcoming geometric limitations \cite{liu2022multiobjective}\cite{meng2021three}. Consequently, UAVs are poised to play a pivotal role in the IoT vision \cite{motlagh2016low}. UAVs, acting as mobile-edge nodes in MEC systems, are gaining significant attention due to their high flexibility and manoeuvrability, capable of providing high-speed, wide-coverage, and low-latency communication services when combined with wireless communication systems \cite{liu2022multi}. Multi-UAV-assisted MEC systems offer unique advantages, including dynamic trajectory planning based on GDs' real-time locations and tasks, contributing to energy savings and reduced latency. Furthermore, UAVs' variable heights increase the probability of establishing LoS links with GDs, strengthening and expanding UAV coverage \cite{xu2021uav}. 

Computation offloading enhances QoS by redirecting tasks to nearby MEC servers. This process is intrinsically linked to effective task scheduling and load balancing. In work \cite{8833501}, the authors presented a two-tier optimization approach to optimize UAV deployment and task scheduling concurrently. The upper layer focused on optimizing UAV deployment, while the lower layer handled task scheduling based on the designated UAV deployment. Yang et al. \cite{8981986} successfully achieved multi-UAV load balancing, ensuring coverage constraints and meeting QoS for IoT nodes. Furthermore, for task scheduling within specific UAVs, a DRL algorithm was developed to enhance task execution efficiency for each UAV. In the realm of communication and computation optimization, energy consumption is a primary focus. Zhang et al. \cite{8877759} addressed this by optimizing bit allocation, time slot scheduling, power allocation, and UAV trajectory design to minimize overall energy consumption, encompassing communication, computation, and UAV flight.

Simultaneously, resource allocation plays a vital role in efficient resource utilization among GDs to prevent resource wastage. In MEC systems, resource allocation closely interacts with computation offloading. Seid et al. \cite{9366889} introduced a model-free DRL-based collaborative resource allocation and computation offloading scheme within an A2G network. Each UAV CH acted as an agent, autonomously allocating resources to Edge IoT (EIoT) devices in a decentralized manner. Yu et al.\cite{yu2020joint} proposed an innovative UAV-enabled MEC system where UAVs and edge clouds (ECs) collaborated to provide MEC services for IoT devices. This proposed system aimed to minimize the weighted sum of service latency and UAV energy consumption for all IoT devices through joint optimization of UAV location, communication, computing resource allocation, and task-splitting decisions. To address the diverse QoS requirements, Peng and Shen \cite{peng2020multi} efficiently employed a multi-agent deep deterministic policy gradient (MADDPG) method to swiftly make vehicle association and resource allocation decisions during the online execution phase. In another approach, Nie et al. \cite{nie2021semi} optimized resource allocation, user association, and power control in a MEC system involving multiple UAVs, introducing a multiagent Federated Reinforcement Learning $($FRL$)$ algorithm while ensuring GDs' privacy protection.

\subsection{ UAV as Wireless Power Transfer (WPT) Source}

WPT is being integrated into wireless communication and computation networks \cite{mao2017survey, mach2017mobile}, enabling a variety of unique applications. These include Simultaneous Wireless Information and Power Transfer (SWIPT) \cite{clerckx2018fundamentals}, Wireless Powered Communication Networks (WPCN) \cite{bi2016wireless}, and Wireless Powered Mobile Edge Computing (MEC) \cite{wang2020optimal}.
The efficiency of energy transfer from Energy Transmitters (ETs) to distributed wireless devices is the primary problem in RF-based WPT systems. First, because of the significant RF signal propagation loss over distance, energy transfer efficiency suffers greatly when devices are positioned far away from the ETs. Second, when several devices are dispersed at different places, the devices close to an ET will harvest substantially more energy than those far away from all ETs, resulting in a major near-far fairness issue. The near-far dilemma may become much more severe in (WPCN) and wireless-powered MEC applications, as far-apart devices with less gathered energy must consume more energy for data transfer to provide the same QoS for nearby devices.

Various approaches have been proposed to address the challenges of efficient energy transfer. These include techniques like transmitting energy beamforming, optimizing energy waveforms, adaptive power control, and strategic deployment planning for ETs. Despite advancements in these areas, achieving widespread coverage for WPT often necessitates deploying ETs densely in fixed locations. However, this approach comes with high deployment and maintenance costs, limiting its practical usability and widespread implementation. The progress in UAV-enabled wireless communications has sparked interest in UAV-enabled WPT as a promising solution to address the technical challenges in WPT. Unlike traditional fixed-location Energy Transmitters (ETs), low-altitude UAVs can function as aerial ETs, offering flexible mobility to charge nearby low-power devices efficiently. UAVs benefit from optimal LoS channels with GDs \cite{al2017modeling} and can dynamically adjust their flight paths to reduce transmission distances to GDs based on real-time locations. This adaptive approach significantly improves energy transfer efficiency to all GDs and ensures fairness in performance. UAV-enabled WPT is particularly well-suited for extensive wireless networks with numerous widely dispersed GDs, where the conventional method of densely deploying fixed-location ETs is impractical due to high costs and feasibility challenges. A fundamental problem in increasing energy transfer in UAV-enabled WPT is effectively planning UAV trajectories. Positioning the UAV close to the GD improves energy transfer in a basic scenario with one UAV and one GD. When several GDs are deployed across known locations, striking a balance in energy transmission among them within a finite charging time presents a significant issue. The proximity of the UAV to a GD for efficient charging may result in suboptimal charging for other GDs, resulting in a fundamental tradeoff. The complexity of joint multi-UAV trajectory design increases as this challenge is scaled to a network with many UAVs jointly charging numerous GDs. \cite{xie2018throughput} investigated a network for WPT using UAVs. The goal was to maximize system throughput by optimizing trajectory and wireless resource allocation. \cite{xu2018uav} investigated how UAV mobility in a WPT network can improve the amount of energy delivered to ground-based energy receivers within a given charging interval. \cite{ye2020optimization} used a UAV-assisted full-duplex wireless-powered IoT model with three optimization challenges: maximize total throughput, minimize overall time, and reduce overall energy consumption. However, their efforts were mostly focused on traditional performance measurements. Furthermore, only a few works considered  RL approaches to enable data gathering within Smart Agriculture-aided IoT (SAG-PIoT) to improve the AoI.

% \section{Future Trends and Open Research Questions}
% \label{sec:future-trends}
% \subsection{Predictions for the Future}
% \subsection{Ongoing Research Trends}
% \subsection{Unresolved Research Questions}
\section{Future Challenges and Solutions}
\label{sec:challenges-solutions}
This section examines the significant forthcoming issues and their potential resolutions in the domain of UAV-assisted IoT, encompassing both interoperability and interference concerns, while leveraging the capabilities of AI to achieve smooth integration and enhance network performance.

% \color{blue}
\subsection{Interoperability}

In IoT networks assisted by UAVs, interoperability is a complex and multifaceted challenge that has a significant influence on the efficiency of communication between IoT devices and UAVs. This difficulty is a result of the coexistence of diverse communication technologies, each of which has its own distinct characteristics and limitations. IoT devices frequently employ low-power, short-range technologies such as 6TiSCH or LoRaWAN for their energy efficiency and suitability for local data transmission. UAVs, on the other hand, rely on a variety of technologies, such as cellular, Wi-Fi, and satellite systems, chosen to facilitate long-range and adaptable communication. When attempting to establish direct communication connections between IoT devices and UAVs, the disparities between their hardware and transmission ranges present formidable obstacles.

In addition, the diversity of communication protocols complicates the interoperability challenge. Typically, IoT devices adhere to standard communication protocols such as MQTT or CoAP, which are widely acknowledged and facilitate seamless data exchange. In contrast, UAVs may utilise custom or proprietary protocols chosen for their compatibility with particular applications and hardware requirements \cite{9869817}. These variations in protocol selection result in diverse message formatting and processing, which makes it difficult to interpret and process data effectively. Even when UAVs and IoT devices employ the same communication protocol, data encoding formats, message structures, and payload syntax may differ. These differences can result in misinterpretation of data, which may cause transmission errors or delays. Moreover, semantic differences resulting from inconsistent terminology, data models, and ontologies exacerbate the interoperability problem. These disparities in data comprehension and interpretation can hinder effective communication between IoT and UAV devices. This misalignment in semantics complicates the exchange of information, which is crucial to the success of IoT applications facilitated by UAVs \cite{sarkar2023artificial}.

It is essential to implement standardised communication protocols, data formats, and semantic models to address these interoperability issues. These standards provide a common basis for data exchange, ensuring that IoT devices, UAVs, and any AI systems can communicate without interruption. In addition, gateways that facilitate translation between different technologies and protocols can help bridge the interoperability gap. These gateways serve as intermediaries, facilitating effective communication between IoT devices and UAVs by converting between various communication technologies and protocols. These gateway devices may also include AI components that facilitate data pre-processing, making the data more readily exploitable by AI algorithms. Middleware solutions designed for semantic harmonisation are crucial for fostering interoperability. By mapping data from various sources to a common ontology, these middleware systems reconcile semantic differences. This ensures that AI systems can accurately interpret data regardless of its origin, thereby enhancing the quality of insights generated by AI algorithms \cite{9869817, sarkar2023artificial}.

In conclusion, the interoperability challenges in UAV-assisted IoT networks are complex and multifaceted, impacting numerous communication aspects. Standardized communication protocols, gateway devices, and semantic middleware solutions are essential to overcoming these obstacles. By assuring reliable and efficient data collection and exchange, UAVs can utilize the full potential of AI for enhanced autonomous decision-making and operations, thereby enhancing the capabilities and efficiency of UAV-assisted IoT applications.
\subsection{Interference in massive UAV-IoT Networks}

In the rapidly evolving landscape of massive UAV-IoT networks, a prominent challenge on the horizon is the issue of interference. As IoT devices proliferate within these networks, each possessing diverse data throughput requirements, mobility needs, and a strong demand for spectral efficiency, the deployment of numerous UAVs becomes imperative to optimize network performance. This escalation in interference is a direct consequence of the high density of UAVs operating in close proximity. The surge in interference has a profound and adverse impact on the overall network performance and data transmission capacity. Adding to the complexity, the operational paradigm in massive UAV-IoT networks is characterized by a shared wireless channel where UAVs exchange information with various entities within the network. This operational configuration results in intricate and often unavoidable wireless interference \cite{9869817}. This interference can be categorized into two primary areas:

\begin{enumerate}
    \item \textbf{Interferences stemming from the coexistence of UAVs and terrestrial macro cells} -- as UAVs coexist with terrestrial cellular networks, challenges related to interference emerge. Differences in transmission powers and communication protocols between UAVs and ground-based cellular infrastructure can result in signal conflicts and reduce overall network efficiency.
    \item \textbf{Interferences arising when the coverage areas of individual UAVs overlap} -- in dense UAV-IoT networks, the coverage areas of individual UAVs are likely to overlap. This scenario can lead to interference as multiple UAVs attempt to serve the same area or IoT devices simultaneously. This results in signal degradation and a reduction in data throughput.
\end{enumerate}

Mitigating interference is of paramount importance, as it holds the potential to significantly enhance the network's throughput, reduce energy consumption, and ensure consistent data delivery. To address this challenge effectively, innovative learning techniques should be developed to promote a high degree of autonomy among UAVs, minimizing the need for extensive information sharing while encouraging self-organization among these vehicles. Additionally, distributed schemes can help alleviate the computational overhead, leading to a reduction in information exchange and, consequently, reducing interference \cite{9324909}. The growing number of IoT devices, particularly in smart city and peripheral technology scenarios, exacerbates the interference dilemma. When these devices are incorporated into large-scale deployments in which UAVs play a central role in data collection and communication, it becomes imperative to employ interference mitigation techniques. The indispensability of these techniques for navigating the complex terrain of vast UAV-IoT networks stems from the fact that they not only aid in mitigating interference but also contribute to network efficiency. In the face of these challenges, integrating AI technologies is poised to play a pivotal role in addressing interference issues. AI can enable UAVs to adapt dynamically to changing interference patterns, optimize their communication strategies, and reduce the overall impact of interference on network performance. The use of AI-driven solutions can lead to more resilient and efficient massive UAV-IoT networks, where interference challenges are met with intelligent responses, ultimately ensuring the continued growth and success of these networks.

\color{black}
\section{Conclusion}

In this article, we provided a comprehensive overview of the operational modes of UAVs, which encompassed static, mobile, and hybrid configurations, along with their associated performance metrics. %Our focus was on the reliable and efficient data collection strategies. 
We explored topics, including data accuracy, network connectivity, and network security protocols, emphasizing their critical role in securing UAV networks against emerging threats and ensuring physical layer security. Furthermore, we showcased advanced security solutions, such as IRS, NOMA, and multi-antenna systems. We also offered insights into effective data-gathering strategies, which covered trajectory optimization, data clustering, aggregation methods, UAV swarm formation, and the integration of cutting-edge optimization algorithms like Reinforcement Learning and Generative Adversarial Networks. Additionally, two practical use cases for UAVs as a service were presented, and we discussed the future challenges in achieving reliable and efficient data collection for UAV-assisted IoT networks. %The information presented in this article laid the foundation for enhancing the capabilities of UAVs in the realm of data collection for IoT networks, ensuring they continued to play a pivotal role in the evolving landscape of modern technology.

% \textcolor{red}{Write the conclusion and also re-write the blue text. I have generated using GPT}
\label{sec:conclusion}
% \subsection{Summary of Key Findings}
% \subsection{Importance of Data Collection}
% \subsection{Call to Action}

%\section*{Acknowledgments}

\bibliographystyle{IEEEtran}
\bibliography{ref}
  
\end{document}